\documentclass[sigconf, screen]{acmart}
\PassOptionsToPackage{table,xcdraw}{xcolor}
\AtBeginDocument{%
  }

\copyrightyear{2025}
\acmYear{2025}
\setcopyright{acmlicensed}
\acmConference[CHI'25]{CHI Conference on Human Factors in Computing Systems}{April 26-May 1, 2025}{Yokohama, Japan}
\acmBooktitle{CHI Conference on Human Factors in Computing Systems (CHI'25), April 26-May 1, 2025, Yokohama, Japan}\acmDOI{10.1145/3706598.3713449} \acmISBN{979-8-4007-1394-1/25/04}

\acmSubmissionID{2068}

\graphicspath{{figures/}{pictures/}{images/}{./}} % where to search for the images

\usepackage{amsmath}
\usepackage{algorithm}

\usepackage{color}
\usepackage{bbding}

\usepackage{enumitem}
\setlist{noitemsep,parsep=0pt,partopsep=0pt, leftmargin=10pt} 
\usepackage{listings}
\usepackage{multirow}
\usepackage{colortbl}
\newcommand{\eg}{{\it e.g.,\ }}
\newcommand{\etal}{{\it et al.\ }}
\newcommand{\etc}{{\it etc.}}
\newcommand{\ie}{{\it i.e.,\ }}
\definecolor{tablerowcolor}{rgb}{0.667,0.667,0.667 }
\definecolor{tablerowcolor2}{rgb}{0,0,0}
\definecolor{visual}{HTML}{e8efd9}
\definecolor{motion}{HTML}{fde7d5}
\definecolor{narrative}{HTML}{e2dce9}
\definecolor{audio}{HTML}{d6ebf2}
\definecolor{bluecrayola}{rgb}{0.12,0.46,1.0}

\newcommand{\visual}[1]{\colorbox{visual}{#1}}
\newcommand{\motion}[1]{\colorbox{motion}{#1}}
\newcommand{\narrative}[1]{\colorbox{narrative}{#1}}
\newcommand{\audio}[1]{\colorbox{audio}{#1}}

\newcommand{\revise}[1]{{\color{black} #1}}
\newcommand{\review}[1]{\textcolor{purple}{}}

\begin{document}
% \title{From Empirical Foundations to Technological Implementation: Understanding Data-Driven Animated Video Storytelling}
% \title{Understanding Animated Data Video Creation Practice}
% \title[]{Lowering the Barrier of Creating Animated Data Videos: Understanding Paradigms with AI Assistance}
% \title[]{Understanding Creation Tools of Animated Data Videos: Interaction Paradigms and AI Assistance}
% \title[]{Component-Centric Understanding of Animated Data Video Creation Tools}
\title{Reflecting on Design Paradigms of Animated Data Video Tools}

\author{Leixian Shen}
\orcid{0000-0003-1084-4912}
\affiliation{%
  \institution{The Hong Kong University of Science and Technology}
  % \streetaddress{P.O. Box 1212}
  \city{Hong Kong SAR}
  % \state{Ohio}
  \country{China}
  % \postcode{43017-6221}
}
\email{lshenaj@connect.ust.hk}

\author{Haotian Li}
\authornote{The work was done when Haotian Li was at HKUST.}
\authornote{Yun Wang and Haotian Li are corresponding authors.}
\orcid{0000-0001-9547-3449}
% \affiliation{%
%   \institution{The Hong Kong University of Science and Technology}
%   \city{Hong Kong SAR}
%   \country{China}
% }
\affiliation{%
  \institution{Microsoft Research Asia}
  \city{Beijing}
  \country{China}
}
\email{haotian.li@microsoft.com}

\author{Yun Wang}
\authornotemark[2]
\orcid{0000-0003-0468-4043}
\affiliation{%
  \institution{Microsoft Research Asia}
  \city{Hong Kong SAR}
  \country{China}
}
\email{wangyun@microsoft.com}

\author{Huamin Qu}
\orcid{0000-0002-3344-9694}
\affiliation{%
  \institution{The Hong Kong University of Science and Technology}
  \city{Hong Kong SAR}
  \country{China}
}
\email{huamin@cse.ust.hk}

\begin{abstract}
%150 words

% Data videos have become increasingly popular for data storytelling, but the creation process remains challenging.
% Numerous creation tools have been developed to streamline the transformation from user inputs into coordinated data video components. However, the complex process of data video creation has not been sufficiently studied. 
% transformation of user inputs into coordinated data video components, 

Animated data videos have gained significant popularity in recent years. However, authoring data videos remains challenging due to the complexity of creating and coordinating diverse components (\eg visualization, animation, audio, \etc). 
Although numerous tools have been developed to streamline the process, there is a lack of comprehensive understanding and reflection of their design paradigms to inform future development.
To address this gap, we propose a framework for understanding data video creation tools along two dimensions: \textit{what data video components to create and coordinate}, including visual, motion, narrative, and audio components, and \textit{how to support the creation and coordination}. 
By applying the framework to analyze 46 existing tools, we summarized key design paradigms of creating and coordinating each component based on the varying work distribution for humans and AI in these tools.
Finally, we share our detailed reflections, highlight gaps from a holistic view, and discuss future directions to address them.
\end{abstract}




% \begin{teaserfigure}
%     \centering
%     \includegraphics[width=0.98\linewidth]{figures/interface.png}
%     \caption{User interface of \toole. Users can first select the text phrases in the narration editor (a) and visual elements from the canvas (b) to form text-visual links. Then they can apply an animation to the visual elements by selecting it from the animation presets in the animation effect panel (c). \tool generates a narration-animation pack and presents it on the timeline (d). 
%     % Finally, users can iteratively preview the video and make new narration-animation packs.
%     }
% \label{fig:user interface}
% \end{teaserfigure}

\maketitle

\section{Introduction}
% Visual data storytelling has become essential across diverse fields such as business, education, design, and marketing, as a story is worth a thousand pictures~\cite{Gershon2001}. The engagement with visual data stories is immense, with platforms like Instagram, YouTube, and TikTok hosting billions of monthly active users interacting with visual content daily. 
Data video is an increasingly popular data storytelling medium across various fields like business, education, and design~\cite{Amini2015, Segel2010}.
% , as a story is worth a thousand pictures~\cite{Gershon2001}.
By integrating visualizations, animations, audio, and other multimedia components, data video can captivate audiences and convey information in an accessible manner~\cite{Borgo2022, Concannon2020}. 
Data videos enjoy significant engagement, with platforms like YouTube and TikTok having billions of monthly active users interacting with such visual content~\cite{active}.
% The engagement with data videos is immense, with platforms like YouTube and TikTok hosting billions of monthly active users interacting with visual content~\cite{active}. 

% One increasingly popular medium of data stories is data video~\cite{Amini2015, Segel2010}. 
Despite the benefits of data videos, data video authoring remains a time-consuming and expertise-intensive process due to the complexity of creating and coordinating diverse data video components~\cite{dataplayer, wonderflow, Shi2021a, Shin2022}. 
Typically, the production of data videos involves data analysts and designers to create static materials and compose stories (\eg visualizations, graphics, text narration, \etc), animators to design dynamic animations and transitions, sound engineers to produce audio narration, music, and sound effects, and multimedia specialists to coordinate all the data video components. Furthermore, it usually involves an iterative workflow to continually refine the produced data video.

% As the demand for data video creation increases, particularly among general users, and with the rapid advancements in AI, numerous tools have been developed to streamline this labor-intensive, expertise-intensive, and creativity-intensive process. 
With the growing interest in data video creation, especially among general users, and the rapid advancements in AI, numerous tools have been developed to streamline the process.
These tools accept diverse user inputs and transform them into a series of coordinated data video components, each involving different roles of humans and AI, covering various design paradigms.
For example, in animation creation within data videos, early efforts tried to reduce users' workload through optimizing manual interactions, such as encapsulating low-level functions as interactive widgets~\cite{Lu2020b} and offering design templates~\cite{Amini2017}. The emergence of AI techniques further automated the process, allowing the recommendation of animation templates~\cite{Shi2021a, dataplayer} and automatic transition generation between keyframes~\cite{Thompson2021, Lee}. Meanwhile, the introduction of AI brought new challenges: humans need to both interact with AI and their creative product, leading to the research on new design paradigms~\cite{DataParticles2023, wonderflow}.
% \textbf{[placeholder]}(For example, manual programming and WIMP interfaces allow users to exert full control over the creation process, albeit with a high level of complexity. AI advancements have made it possible to automate tasks with clear patterns, enabling the intelligent extraction of user intent from various interaction modes—such as natural language, sketches, and examples—for subsequent tasks. The advent of large language models has further revolutionized these paradigms, enhancing the capabilities and accessibility of data video creation tools.)

\review{Q1}\revise{Recent research has analyzed storytelling tools and specific storytelling form design patterns. However, they are coarse-grained (\eg based on story mediums~\cite{Segel2010, Zhao2023a} or AI roles within the system~\cite{Li2023c, Chen2022}), primarily empirical~\cite{Hao2024, Bach2018, Bach2023, Lan2023}, and not tailored for data videos. 
Data videos integrate complex components, each presenting unique challenges in creation and coordination.
So existing works can not support a fine-grained analysis of technical implementations, especially at the component and coordination level, which is vital to inform the design of future tools.}
% They miss sufficient understanding and reflection of the design paradigms for data video creation, which is vital to inform the design of future tools.
% has not been sufficiently studied. complex data video creation workflow 
% There is a lack of systematic reviews on the methodologies for creating data videos. 
To address this gap, \review{Q4}\revise{we collected and analyzed 46 tools that produce data videos played on 2D screens.} We proposed a framework that organizes these tools along two dimensions: \textit{what data video components to create and coordinate}, 
% (\ie the involved types of components and their coordination) 
and \textit{how to support the creation and coordination}.
% (\ie how user inputs are transformed into coordinated components).
% component complexity (i.e., the involved types of components and their coordination) and component transformation mode (i.e., how user inputs are transformed into coordinated components).
Specifically, we first analyzed data video structures and decomposed data video components, classifying them into four types: visual components, motion components, narrative, and audio components. 
% Greater diversity in data video components enhances expressive potential but also increases authoring complexity.
A greater diversity of components in a data video can enhance its expressive potential.
% We argue that the more diverse the components in a data video, the greater its expressive potential.
Therefore, in the first dimension, we further categorized data videos into three levels of expressivity based on component combinations: animation unit (comprising only visual and motion components), animated narrative (which additionally includes narrative), and audio-enriched data video (which further integrates audio components). 
Each category requires careful consideration of the specific components and their coordination.
For the second dimension, these tools encompass varying work distribution for humans and AI to facilitate the creation and coordination of each component. Therefore, we categorized the transformation process from user input into the final coordinated data video components into four modes: original (without transformations), human-led, mixed-initiative, and AI-led. 
% Using this framework, we conducted an extensive review of existing research on data video creation tools. 
% Within this framework, we analyzed the processes of generating and coordinating each component and identified key design paradigms within each transformation mode.
% as more data video components are involved,
Within this framework, we analyze how existing tools assist users in creating and coordinating each component with less effort,
% based on the varying roles of humans and AI, 
identifying key design paradigms under each transformation mode.
We also present our detailed reflections on the these design paradigms. Finally, we take a holistic view of data video production, highlighting gaps at each stage and discussing potential directions to address them.
% Within this framework, we analyzed how existing tools assist users in generating and coordinating each component with less effort based on the different work distributions of humans and AI, identifying key design paradigms within each transformation mode.
% Finally, we reported our findings, provided design suggestions, and discussed future research directions. 
Through this paper, we aim to enhance the understanding of current data video creation practices and promote the development of more effective and efficient tools and methodologies.
Overall, our main contributions are:
\begin{itemize}
\item 
We propose a framework for understanding data video creation tools along two dimensions: \textit{what data video components to create and coordinate}, and \textit{how to support the creation and coordination of these components}.
\item 
We provide a thorough analysis of the existing 46 animated data video tools within the framework, summarizing the key design paradigms for each component and coordination with different roles of humans and AI.
\item
We present our reflections in detail, highlight remaining gaps from a holistic view, and discuss potential future directions to bridge the gaps.
\end{itemize}

% \begin{compactitem}
% \item We introduce a framework for data videos creation based on expressivity and learnability, defining several distinct genres of data videos.
% \item  We provide a thorough survey of existing research, summarizing the current state of empirical studies and technical implementations in the field.
% \item We analyze the existing research gap between these two aspects and discuss potential future directions to enhance the integration of empirical insights with technical advancements.
% \end{compactitem}

\section{Related Work}\label{sec:related}
A data story is composed of a set of data-driven story pieces, which are typically visualized and presented in a meaningful order to communicate the author's intended message \cite{Lee2015, Riche2018}. 
Segel and Heer~\cite{Segel2010} identified seven genres of narrative visualization: magazine style, annotated chart, partitioned poster, flow chart, comic strip, slide show, and data video. 
Tong \etal~\cite{Tong2018} classified the storytelling literature based on key aspects of storytelling in visualization: the subjects involved, the methods used to convey stories, and the underlying purpose of storytelling in visualization.
Zhao and Elmqvist~\cite{Zhao2023a} proposed a taxonomy for data storytelling, focusing on five dimensions: audience cardinality, space and time, media components, data components, and viewing sequence.

For data story authoring, 
Li \etal~\cite{Li2023c} analyzed existing data storytelling tools from the perspective of human-AI collaboration, outlining four key stages in the storytelling process—analysis, planning, implementation, and communication—and identifying the roles of humans and AI as creators, assistants, optimizers, and evaluators. 
Chen \etal~\cite{Chen2022} developed a framework that maps seven types of narrative visualization against four levels of automation, analyzing how automation affects the storytelling process.
Ren \etal~\cite{Ren2023} categorized data storytelling tools based on whether the narrator offers a personal perspective on the data, distinguishing between omniscient and limited perspectives. 
Shi \etal~\cite{Shi2023} surveyed \textit{AI for design} to analyze how designers and AI can complement and collaborate with each other, identifying three themes: AI assisting designers, designers assisting AI, and collaboration between the two. 
% Wu \etal \cite{Wu2021d} and Wang \etal \cite{Wang2020c} provided insights into how AI can enhance the visualization process. 
% \haotian{Why existing AI work cannot address our question?}
\review{Q1, Q11}\revise{
However, the research above primarily offers a coarse-grained classification and discussion of existing data storytelling tools (\eg based on story mediums or AI roles within the system) and is not tailored for data videos. 
Data video combines diverse coordinated components (\eg visualization, animation, audio, \etc), and each component and coordination relationship poses different challenges.
As a result, existing frameworks do not support a fine-grained analysis of technical implementations for data videos.
}
% However, the above research is not tailored for data videos, missing sufficient understanding and reflection on core technical aspects of data video creation.

In addition to broader surveys, some research has also explored design patterns for specific data storytelling forms, such as data-driven news articles~\cite{Hao2024}, data comics~\cite{Bach2018}, dashboards~\cite{Bach2023}, affective visualizations~\cite{Lan2023}, timelines~\cite{Brehmer2017}, and character-oriented designs~\cite{Dasu2024, Mittenentzwei2023a}. 
% data videos, as a specific form of data storytelling, possess a complex structure that general surveys cannot fully cover in terms of core technical aspects.
In the context of data videos, numerous studies have empirically examined specific components such as narrative structure~\cite{Lan2021a, Yang2022a, Cao2020a, Wei2024}, cinematic styles~\cite{Xu2023b, Xu2022, Concannon2020, Conlen2023, Bradbury2020}, narrative transitions~\cite{Tang2020}, narration-animation interplay~\cite{Borgo2022}, animation roles and effectiveness~\cite{Amini2018a, Chevalier2016, Shu2020}, affectiveness~\cite{Sakamoto2022, Sallam2022}, visualization in motion~\cite{Yao2022}, and 3D data videos~\cite{Yang2023a}. 
\review{Q1, Q11}\revise{However, these studies are limited to empirical research and do not address technical implementation.

Overall, there is a lack of a systematic survey that reflects existing technological practices in data video creation, summarizing key design paradigms for creating and coordinating fine-grained data video components. 
This is vital for guiding new researchers and informing the development of next-generation design paradigms and tools.
Therefore, this paper will provide an in-depth technical analysis at the component level for data video production.}
% Despite the increasing attention of data videos in information communication and the rapid development of AI, there lacks a comprehensive survey on how to create data videos by designing paradigms and leveraging AI assistance to lower the barrier. 
% This gap is particularly significant in helping researchers understand existing paradigms and the role of AI in inspiring future research. 

\section{Survey Landscape}\label{sec:Landscape}
% \subsection{Scope}
To compile a comprehensive corpus of data videos,
% , we employed a keyword-based search strategy.
we focused on papers from the following key venues: Visualization (VIS), including TVCG, CGF, CGA, IEEE VIS, EuroVis, PacificVis, and AVI; Human-Computer Interaction (HCI), including CHI, UIST, IUI, and CSCW; and Computer Graphics (CG), including SIGGRAPH and MM. 
Specifically, we conducted a keyword search in Engineering Village from 2000 to 2024 (April) based on papers' titles, abstracts, and index terms.
The search keywords were derived from Amini’s definition of data videos~\cite{Amini2015}, which requires that the video content must contain data-backed arguments to support its central message, include at least one data visualization, and use a narrative format. 
The keywords used were: (data) AND (visualization OR visual OR story OR storytelling OR chart) AND (animation OR video OR motion OR visual cue).
This resulted in 688 papers, which we supplemented by reviewing references to capture a broad spectrum of relevant research.
% Additionally, we reviewed references from the filtered papers to include other representative related works in other venues (\eg arXiv). This allowed us to capture a broad spectrum of relevant research while also identifying key papers through citation networks.

\begin{figure}[t]
  \centering
    \includegraphics[width=\linewidth]{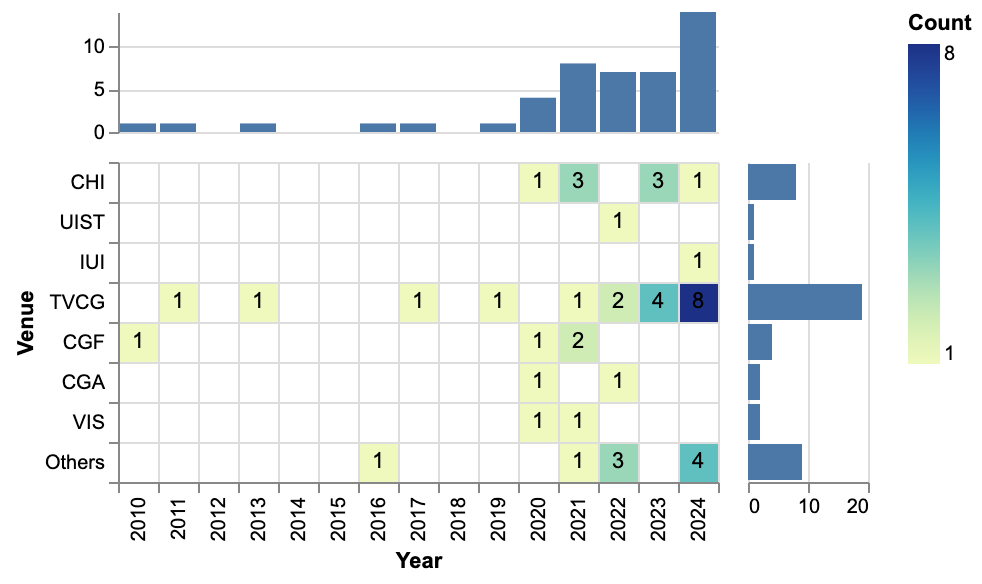}
     % \vspace{-10px}
    \caption{Overview of papers about data video creation tools across publication years and venues.}
\label{fig:overview}
  % \vspace{-10px}
\end{figure}

When reviewing the papers, we applied the following criteria to select the corpus:
First, the works should serve the purpose of data storytelling, excluding those for data analysis (\eg scientific visualization animations~\cite{Morth2023}). 
Second, the works should be data-driven, excluding general videos that merely tell a narrative without data~\cite{Chi2022}. We also excluded text or word cloud animations~\cite{Xie2023a, Xie2023}.
% (\eg Wakey-Wakey~\cite{Xie2023a} and Emordle~\cite{Xie2023}).
% Finally, we temporarily excluded 3D data videos related to virtual reality~\cite{Hong}, as their production paradigms differ significantly from 2D data videos. 
\review{Q4}\revise{Third, we excluded 3D data videos associated with virtual reality (VR) devices~\cite{Yang2023a} and concentrated solely on 2D screens, due to the significantly different production paradigms between them.
For example, generating videos in VR devices necessitates complex 3D object modeling, with the video primarily dependent on recordings made with VR headsets' camera moves. Our focus, however, centers on users interacting on 2D screens to create and coordinate various components.}
% However, we will discuss this category further in the discussion section.
For each paper, we first read the title and abstract to make a preliminary judgment. If the judgment was inconclusive, we examined the full text for a final conclusion. During this process, we also categorized the papers into empirical research (45 papers) and creation tools (46 papers). 
This paper focuses primarily on creation tools, with empirical research discussed only incidentally. 
An overview of the creation tools is shown in Fig.~\ref{fig:overview}, which illustrates the growing interest in data video creation research.

% \subsection{Paper Coding}

\section{Framework}
% We utilize the two primary dimensions to model the creation of data videos: expressivity (\ie what data videos can be created) and learnability (\ie How difficult it is to create such a data video).
In this section, we will first decompose the data video components, explaining the function and role of each part while categorizing them into four types. 
Next, we will introduce a two-dimensional framework for understanding data video creation tools: \textit{what data video components to create and coordinate}, and \textit{how to support the creation and coordination}.
% This section will introduce the two-dimension framework for understanding data video creation tools. This section will first decompose the components of data videos, explain each part's function and role, and categorize them into four main groups. 
% Regarding expressivity, we classify data videos into three levels, discussing the required components and how they coordinate at each level. For learnability, we analyze the roles of humans and AI in transforming diverse inputs into various output components.

% in general authoring systems~\cite{Satyanarayan2019a}, 

% \subsection{Expressivity}
% Expressivity reflects the capability of a tool to create various types of data videos. 
% A data video comprises a series of interconnected components, each playing a distinct role but collectively enhancing expressivity. 
% The use and coordination of these components result in data videos with varying levels of expressivity. 
% This section will first decompose the components of data videos, explaining each part's function and role, and categorize them into four main types groups. 
% Subsequently, it will classify data videos into three levels of expressivity based on different combinations of these components and provides a detailed analysis of the additional components and coordination required to achieve higher levels of expressivity.

\subsection{Decomposition of Data Video Components} \label{sec:components}
Data
We decompose data videos into different components (see Fig.~\ref{fig:components}) and categorize them into four groups: visual components, motion components, narrative, and audio components. We use a real-world data video about ``Dubai's Plan to Outlive Oil'' to introduce the components~\cite{Dubai}.
% \footnote{Data Video about Dubai's Plan to Outlive Oil. \url{https://www.youtube.com/watch?v=-dsM8zPGqa0&t=543s}} 

% \vspace{-3px}
\subsubsection{\visual{Visual Components}}\label{sec:visualcomponents}
\begin{figure*}[t]
  \centering
    \includegraphics[width=\linewidth]{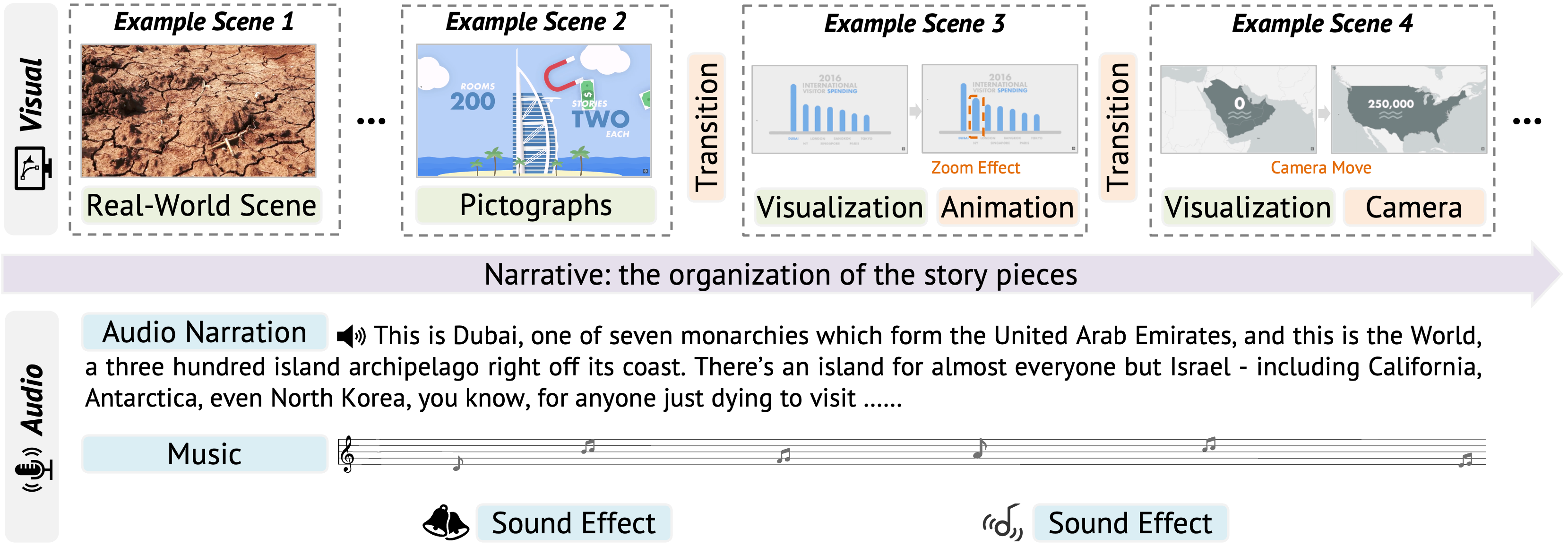}
    % \vspace{-10px}
    \caption{Illustrations of data video components along a timeline from left to right, each serving a distinct role. They are divided into four categories: \visual{visual components}, \motion{motion components}, \narrative{narrative}, and \audio{audio components}.}
\label{fig:components}
  % \vspace{-10px}
\end{figure*}

Visual components are the backbone of data videos.
% , conveying data and context while engaging the audience visually. 
They include various visual representations that translate data into understandable formats, as well as non-data visuals that provide contextual information.
% \vspace{-3px}
\begin{itemize}
\item 
\textit{Data Visualization:} 
Visualizations transform abstract numbers into intuitive visual formats, interpreting data insights in data videos. For instance, the bar chart in Fig.~\ref{fig:components} (Example Scene 3) illustrates international visitor spending.
% Data Visualizations are essential elements in data videos~\cite{Amini2015}, crucial for interpreting data insights. They transform abstract numbers into intuitive visual formats, allowing viewers to quickly grasp trends, patterns, and insights. For instance, the bar chart in Fig.~\ref{fig:components} (Example Scene 3) illustrates international visitor spending.

\item 
\textit{Real-World Scene:} 
Integrating real-world images or footage provides context that grounds abstract data in reality and adds an emotional dimension. 
% This approach not only makes the data more relatable but also adds an emotional dimension, helping viewers to connect with the information on a personal level. 
For example, Fig.~\ref{fig:components} (Example Scene 1) uses a real-world image to describe that large portions of the country's land are covered in a salty crust.
\item 
\textit{Pictograph:} 
Pictographs use virtual graphics and images to illustrate complex concepts~\cite{Jahanlou2022}. They are flexible and play a supplementary role in data videos to clarify information and enrich narratives.
% Compared to data visualizations, pictographs offer high flexibility and play a supplementary role in data videos by clarifying information and enriching the narrative, especially for concepts that might be hard for data visualizations and real-world scenes to express. 
For instance, Fig.~\ref{fig:components} (Example Scene 2) uses pictographs to depict the Burj AI Arab hotel, illustrating its room condition and revenue potential.

% , making information more understandable. money-making property.
\end{itemize}
% % \vspace{-3px}

\begin{figure*}[t]
  \centering
    \includegraphics[width=\linewidth]{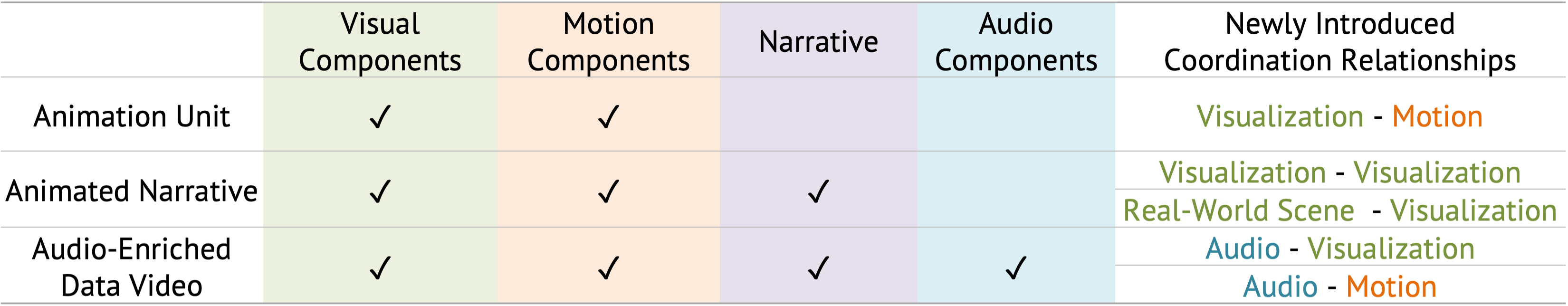}
    % \vspace{-10px}
    \caption{Data video with three levels of expressivity. Greater diversity in data video components enhances expressive potential but also increases authoring complexity, due to involving more components and introducing new coordination relationships.
    }
\label{fig:expressivity}
  % \vspace{-10px}
\end{figure*}

% \vspace{-3px}
\subsubsection{\motion{Motion Components}}\label{sec:motioncomponents}
Motion components add dynamism to visual components, making the content more engaging and easier to follow. This includes the in-scene animations and seamless transitions between different scenes.
% \vspace{-3px}
\begin{itemize}
\item
\textit{Animation:} 
Animations catch viewers' attention and guide navigation by showing changes in visual channels~\cite{Chevalier2016}. 
Animations commonly include three behaviors, \ie entrance, emphasis, and exit. Common animation types include fade, grow, float, fly, zoom, change color, \textit{etc.}
For example, the zoom effect on the London bar in Fig.~\ref{fig:components} (Example Scene 3) highlights that international visitors to Dubai spend more than any other city in the world, including London.
\item
\textit{Transition:} 
Transitions between charts or between data-driven charts and non-data graphics in the video help maintain narrative flow and direct viewers' attention~\cite{Tang2020, Heer2007, Kim2019c, Chalbi2020}. 
While similar to animations, smooth transitions specifically highlight key changes in different video segments, guiding viewers as one scene shifts to another.

\item
\textit{Camera:} 
Camera movements (\eg panning, zooming, and tilting) guide the viewer's focus. They can emphasize specific data elements or scenes and create an immersive viewing experience. For example, Fig.~\ref{fig:components} (Example Scene 4) shows a camera move from Dubai to America on the world map.
% and contribute to the narrative structure of the video. These movements
\end{itemize}
% \vspace{-3px}

\subsubsection{\narrative{Narrative}}\label{sec:narrativecomponents}
% In addition to visual, motion, and audio components, 
Narrative is essential in data videos~\cite{Amini2015}, weaving together various components into a cohesive story. 
It often involves the organization of story pieces presented through visual components~\cite{Lan2021a, Cao2020a, Riche2018, Schroder2023}, providing structure and context that enhance comprehension and engagement.

\subsubsection{\audio{Audio components}}\label{sec:audiocomponents}
Audio components complement the visual and motion components, providing an additional layer of information and emotional engagement~\cite{clark_dual_1991}. They include various forms of voiceover and sound.
% that enhance the narrative and emphasize key points.
% \vspace{-3px}
\begin{itemize}
\item 
\textit{Audio Narration:} 
Human cognition allows for the simultaneous perception of both visual and audio information~\cite{Obie2019}. Audio narration is the most commonly used audio components that provides a verbal explanation of the data, adding context and depth beyond visuals. 
% It also guides the audience through the data, enhancing users’ comprehension and memorability of visualizations. 
\item 
\textit{Music:} 
Background music typically runs throughout the video and sets the tone and mood.
% , influencing the viewer's emotional engagement. 
It can make the presentation more memorable and enjoyable, though it must be balanced to avoid overshadowing the data itself~\cite{Xu2022, Xu2023b, Concannon2020, Tang2022}.
\item
\textit{Sound Effect:} 
Sound effects are used only at specific moments to emphasize specific data points or transitions. They provide auditory cues that reinforce visual changes and enhance the viewer's focus~\cite{Xu2022,Xu2023b}.
% by making the data presentation more interactive
\end{itemize}
% \vspace{-3px}

\begin{figure*}[t]
  \centering
    \includegraphics[width=\linewidth]{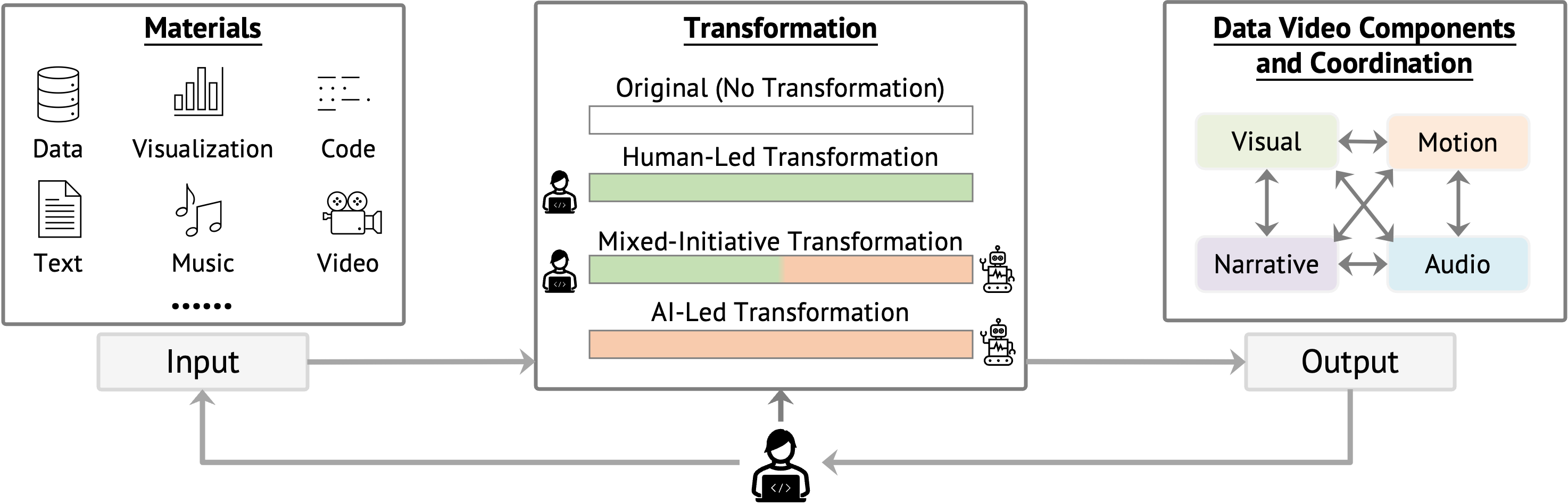}
    % \vspace{-10px}
    \caption{The roles of humans and AI in transforming diverse user inputs into coordinated data video components.
    }
\label{fig:learnability}
  % \vspace{-10px}
\end{figure*}

% \subsubsection{Coordination of Data Video Components}
% \subsection{Expressivity}\label{sec:expressivity}
\subsection{What Data Video Components to Create and Coordinate}\label{sec:expressivity}
% Expressivity reflects the capability of a tool to create various types of data videos. 
% \review{Q9}\revise{A tool's expressivity reflects what data videos can be created using it~\cite{Satyanarayan2019a}.}
% and collectively enhancing expressivity. 
% The way these components are used and coordinated affects the expressivity of the video. 
% Greater diversity in data video components enhances expressive potential, but also increases authoring complexity.
\review{Q9}\revise{
A tool's expressivity reflects what data videos can be created using it~\cite{Satyanarayan2019a}.
Tools that support a wider range of data video components offer an expanded design space, leading to increased expressive potential of final outputs, yet also increasing authoring complexity.
}
\review{Q11}\revise{Data video comprises a series of interconnected components, each fulfilling specific roles, and various tools accommodate different component combinations tailored to distinct tasks and challenges.
To explore fine-grained technical implementations of these data video tools, considering the intricate nature of data videos with complex interconnected components, we first utilize component combinations to classify data videos into three expressivity levels (Fig.~\ref{fig:expressivity}) and will analyze each component and coordination in detail.}
% The broader the range of data video components supported by tools, the larger the design space for creating data videos, leading to increased expressive potential, yet also increasing authoring complexity.
% More diverse data video components that tools support provide a wider design space for creating expressive data videos, but also increases authoring complexity.
% Higher expressivity involve more sophisticated use of components and a higher degree of integration, resulting in more impactful and versatile data videos.
% The use and coordination of these components result in data videos with varying levels of expressivity. We categorize data videos into three levels of expressivity based on the involved components and their required coordination relationships to achieve higher levels of expressivity, as illustrated in Fig.~\ref{fig:expressivity}.

% In this framework, we classify data videos into three levels of expressivity based on different combinations of these components and provides a detailed analysis of the additional components and coordination required to achieve higher levels of expressivity.
% The various components of data videos each play an important role, but they must be effectively coordinated to create an expressive final product. 

\subsubsection{\textbf{Animation Unit}} 
Animation unit comprises only visual and motion components, serving as the minimal unit of visual motion, akin to a single animation in PowerPoint. This necessitates consideration of the relationship between visualizations and motions:
% Animation Unit includes essential visual components that are animated to convey data. This category focuses primarily on visualizations enhanced through motion. The coordination relationship here is primarily between visualization and animation, where the animations serve to highlight key data points and facilitate viewer engagement. 
% Animation Unit focuses primarily on visual and motion components. It leverages animations to enhance data visualizations, making the information more engaging and easier to interpret.
% \vspace{-3px}
\begin{itemize}
\item 
\textit{\visual{Visualization}---\motion{Motion}:} 
This coordination pertains to how static data visualizations are animated to draw attention to specific elements or patterns. Animation effects must align with the shape and semantics of the visuals; for example, a bar chart is suited for a growing effect, while a pie chart benefits from a wheel effect. Zoom effects are used to emphasize details, like in Fig.~\ref{fig:components} (Example Scene 3), and wipe effects effectively demonstrate trends.
% For example, a bar chart may use a zoom effect to emphasize significant spending by international visitors in Dubai. 
\end{itemize}
% \vspace{-3px}

\subsubsection{\textbf{Animated Narrative}} 
Animated narrative expands upon the animation unit by integrating narratives, similar to a complete slide animation sequence or a series of slides in PowerPoint that form a cohesive story. In addition to the visualization---animation relationship, it considers the organization of visual components and additionally introduces:
% The Animated Narrative expands upon the Animation Unit by incorporating both visual and narrative elements. This category utilizes animations alongside structured storytelling to guide the viewer through the data. The coordination relationships involved include visualization to visualization (linking different data points) and visualization to real-world scenes (contextualizing data within familiar frameworks), enhancing the narrative's effectiveness.
% The Animated Narrative expands upon the Animation Unit by integrating narrative elements. This category utilizes both visualizations and storytelling techniques to guide the viewer through complex information.
% \vspace{-3px}
\begin{itemize}
\item 
\textit{\visual{Visualization}---\visual{Visualization}:} 
Data narratives are conveyed through the sequence of data-driven visualizations~\cite{Schroder2023}. This involves considering the temporal and causal relationships among story elements and ensuring seamless transitions between multiple keyframes to facilitate information comprehension~\cite{Kim2017b}.

\item 
\textit{\visual{Real-World Scene}---\visual{Visualization}:} 
Integrating real-world scenes contextualizes abstract data. So the coordination between non-data scenes and data-driven visualizations is also important for narrative coherence. This involves the organic embedding of visualizations within real-world contexts and seamless transitions between the contextual information and the data narrative.
% Integrating real-world scenes contextualizes abstract data, thus enhancing realism and credibility. The coordination between non-data real-world scenes and data-driven visualizations is equally important, involving the organic embedding of visualizations within real-world contexts and seamless transitions between the contextual information provided by real-world scenes and the data narrative.
% For example, showing a map of Dubai alongside data about its economy helps viewers connect the figures to real-life implications, thereby enhancing comprehension.

\end{itemize}
% \vspace{-3px}

\subsubsection{\textbf{Audio-Enriched Data Videos}} 
Audio-enriched data videos build upon the animated narrative by incorporating audio components. This is comparable to a complete PowerPoint presentation video with a presenter’s voiceover. Beyond the previously mentioned coordination, it additionally introduces:
% The Narration-Enriched Data Video represents the highest level of expressivity, integrating visual, motion, and audio components. In this category, narration adds depth and context, enriching the viewer's understanding. The coordination relationships include visualization to narration (providing verbal context for visual data) and animation to narration (aligning animated elements with the spoken narrative), thereby creating a cohesive and immersive experience.
% The Narration-Enriched Data Video represents the highest level of expressivity, incorporating visual, motion, and audio components to create a comprehensive narrative experience.

% \vspace{-3px}
\begin{itemize}
\item 
\textit{\audio{Audio}---\visual{Visualization}:} 
This relationship ensures that verbal explanations or music effects align with visual content, providing context that deepens understanding. For example, as a chart displays spending data, the narration can explain its significance, reinforcing the message.
\item 
\textit{\audio{Audio}---\motion{Motion}:} 
This coordination highlights how motion elements should be synchronized with the spoken narrative or sound effects, reinforcing key narrative points. 
For example, an animation depicting growth trends can coincide with a narrator discussing those trends, accompanied by appropriate background music and sound effects.
% For instance, an animation illustrating growth trends can co-occur with a narrator discussing those trends, with suitable background music and sound effects.
% , creating a more immersive experience.
\end{itemize}
% \vspace{-3px}

We note that this paper focuses only on the coordination relationships closely related to data video production, as described in Sec.~\ref{sec:Landscape}.
% , with the primary components being data visualization, data animation, and data narration. 
Other relationships, such as those between real-world scenes and animations, audio and pictographs, are outside the scope of this discussion, as they are the focus of other research domains.

\begin{table*}[h]
% \small
\centering
\caption{
Overview of existing data video creation tools, categorized into three groups by component combinations as described in Sec.~\ref{sec:expressivity}. 
For the components and coordination relationships outlined in Sec.~\ref{sec:expressivity}, each tool is labeled with its transformation mode (Sec.~\ref{sec:learnability}): Original (\textbf{O}), Human-Led (\textbf{H}), Mixed-Initiative (\textbf{M}), and AI-Led (\textbf{A}). 
% They are categorized by expressivity (Sec.~\ref{sec:expressivity}): animation unit, animated narrative, and audio-enriched data video. Each work is labeled with four transformation modes (Sec.~\ref{sec:learnability}) from user input to the necessary data video components and their coordination: Original (\textbf{O}), Human-Lead (\textbf{H}), Mixed-Initiative (\textbf{M}), and AI-Lead (\textbf{A}).
}
% \vspace{-10px}
\label{tab:tools}
\setlength{\tabcolsep}{3mm}{
\renewcommand\arraystretch{0.7}
\begin{tabular}{cllc*{3}{>{\centering\arraybackslash}p{0.1cm}}*{3}{>{\centering\arraybackslash}p{0.1cm}}*{3}{>{\centering\arraybackslash}p{0.1cm}}}
\textbf{Genre}& 
\textbf{Tool Name} & 
\textbf{Source} & 
\textbf{Year} & 
\rotatebox{50}{\visual{\textbf{Visual}}} & 
\rotatebox{50}{\motion{\textbf{Motion}}} & 
\rotatebox{50}{\textbf{Visualization---Motion}} & 
\rotatebox{50}{\narrative{\textbf{Narrative}}} & 
\rotatebox{50}{\textbf{Visualization---Visualization}} & 
\rotatebox{50}{\textbf{Real-World Scene---Visualization}} & 
\rotatebox{50}{\audio{\textbf{Audio}}} & 
\rotatebox{50}{\textbf{Audio---Visualization}} & 
\rotatebox{50}{\textbf{Audio---Motion}} \\
\midrule
 \multirow{7}{*}{\rotatebox{90}{\color[HTML]{000000}Animation Unit}} 
 & D3~\cite{Bostock2011} & TVCG & 2011 & H & H & H &  &  &  &  &  &  \\
  & Gemini~\cite{Kim2020} & TVCG & 2021 & O & H & A &  &  &  &  &  &  \\
 & Gemini 2~\cite{Kim} & VIS & 2021 & O & A & A &  &  &  &  &  &  \\
 & AniVis~\cite{Li2021c} & arXiv & 2021 & O & H & A &  &  &  &  &  &  \\
 & Lu et al.~\cite{Lu2020a} & TVCG & 2022 & O & A & A &  &  &  &  &  &  \\
 & Epigraphics~\cite{Zhou2024} & CHI & 2024 & M & M & A &  &  &  &  &  &  \\
 & Keyframer~\cite{Tseng2024} & arXiv & 2024 & O & M & M &  &  &  &  &  &  \\
 \bottomrule
 \multirow{23}{*}{\rotatebox{90}{\color[HTML]{000000}Animated Narrative}} 
 & Yu et al.~\cite{Yu2010} & CGF & 2010 & O & A & A & A & A &  &  &  &  \\
 & SketchStory~\cite{Lee2013} & TVCG & 2013 & M & M & A & H & M &  &  &  &  \\
 & Wang et al.~\cite{Wang2016e} & SA & 2016 & A & A & A & A & A &  &  &  &  \\
 & DataClips~\cite{Amini2017} & TVCG & 2017 & M & M & A & H & M &  &  &  &  \\
 & Narvis~\cite{Wang2019a} & TVCG & 2019 & O & H & H & H & H &  &  &  &  \\
 & data2video~\cite{Lu2020b} & CGA & 2020 & M & M & A & A & A &  &  &  &  \\
 & Canis~\cite{Ge2020} & CGF & 2020 & O & H & H & H & H &  &  &  &  \\
 & Vis-Annotator~\cite{Lai2020a} & CHI & 2020 & O & A & A & O & A &  &  &  &  \\
 & Visual Foreshadowing~\cite{Li2020c} & VIS & 2020 & A & M & H & A & A &  &  &  &  \\
 & InfoMotion~\cite{Wang2021d} & CGF & 2021 & O & A & A & A & A &  &  &  &  \\
 & AutoClips~\cite{Shi2021a} & CGF & 2021 & A & A & A & A & A &  &  &  &  \\
 & Cast~\cite{Lee} & CHI & 2021 & O & M & A & H & H &  &  &  &  \\
 & Data Animator~\cite{Thompson2021} & CHI & 2021 & O & M & A & H & H &  &  &  &  \\
 & Datamations~\cite{Pu2021} & CHI & 2021 & A & M & A & A & M &  &  &  &  \\
 & VisCommentator~\cite{Chen2022h} & TVCG & 2022 & M & M & A & O & H & M &  &  &  \\
 & DataParticles~\cite{DataParticles2023} & CHI & 2023 & M & M & A & O & M &  &  &  &  \\
 & GeoCamera~\cite{Li2023b} & CHI & 2023 & O & M & A & H & H &  &  &  &  \\
 & Animated Vega-Lite~\cite{Zong2022} & TVCG & 2023 & O & H & A & H & H &  &  &  &  \\
 & Omnioculars~\cite{Lin2023b} & TVCG & 2023 & A & M & A & O & H & M &  &  &  \\
 & iBall~\cite{Chen2023d} & CHI & 2023 & A & M & A & O & H & M &  &  &  \\
 & SwimFlow~\cite{Yao2024} & TVCG & 2024 & H & M & A & O & H & H &  &  &  \\
 & VisTellAR~\cite{Tong2024} & TVCG & 2024 & H & H & H & H & H & H &  &  &  \\
 & V-Mail~\cite{Nam2024} & TVCG & 2024 & O & M & A & H & H &  &  &  &  \\
 & \review{Q7}\revise{CAST+}~\cite{Shen2024b} & TVCG & 2024 & O & M & A & M & M &  &  &  &  \\
 \bottomrule
  \multirow{13}{*}{\rotatebox{90}{\color[HTML]{000000}Audio-Enriched Data Video}} 
 & Fidyll~\cite{Conlen2022} & arXiv & 2022 & O & M & H & H & H &  & A & M & M \\
 & SmartShots~\cite{Tang2022} & TIIS & 2022 & M & A & A & O & H & M & O & A & M \\
 & Talking Realities~\cite{Latif2022} & CGA & 2022 & O & A & A & A & A &  & A & A & A \\
 & Hall et al.~\cite{Hall2022} & UIST & 2022 & O & M & M & H & H & H & H & H & H \\
 & DataTV~\cite{Zhao2022} & arXiv & 2022 & O & H & H & H & H & H & H & H & H \\
 & Roslingifier~\cite{Shin2022} & TVCG & 2023 & A & A & A & A & A &  & A & M & M \\
 & Sporthesia~\cite{Chen2022c} & TVCG & 2023 & M & M & A & O & H & M & O & M & M \\
 & Data Player~\cite{dataplayer} & TVCG & 2024 & O & A & A & O & A &  & A & A & A \\
 & Live Charts~\cite{Ying2023} & TVCG & 2024 & O & A & A & A & A &  & A & A & A \\
 & VisConductor~\cite{Femi-Gege2024} & ISS & 2024 & O & M & M & H & H & H & H & H & H \\
 & Shi et al.~\cite{Shi2023b} & Pvis & 2024 & O & A & A & O & A &  & O & A & A \\
 & WonderFlow~\cite{wonderflow} & TVCG & 2024 & O & H & A & O & H &  & A & H & H \\
 & AiCommentator~\cite{Andrews2024} & IUI & 2024 & M & M & A & O & H & M & A & M & M \\
 & \review{Q7}\revise{Data Playwright}~\cite{dataplaywright} & TVCG & 2024 & O & M & M & H & M &  & A & M & M \\
 & \review{Q7}\revise{Narrative Player}~\cite{NarrativePlayer} & TVCG & 2025 & A & A & A & O & A &  & A & A & A \\
\hline

\multicolumn{12}{l}{} \\

\hline
 &  & \multicolumn{2}{r}{Original (\textbf{O})} & 26 & 0 & 0 & 13 & 0 & 0 & 3 & 0 & 0 \\
\multicolumn{2}{c}{} & \multicolumn{2}{r}{Human-Led (\textbf{H})} & 3 & 9 & 7 & 15 & 20 & 5 & 3 & 4 & 4 \\
\multicolumn{2}{c}{\multirow{-2}{*}{Statistics}} & \multicolumn{2}{r}{Mixed-Initiative (\textbf{M})} & 9 & 23 & 4 & 1 & 6 & 6 & 0 & 5 & 6 \\
 &  & \multicolumn{2}{r}{AI-Led (\textbf{A})} & 8 & 14 & 35 & 10 & 13 & 0 & 9 & 6 & 5\\
\bottomrule
\end{tabular}}
\end{table*}

% We explored the design paradigm and the role of human and AI in creating data videos with varying levels of expressivity.

\subsection{How to Support Creation and Coordination of Data Video Components}\label{sec:learnability}
% Learnability indicates how difficult it is to create a specific data video~\cite{Grossman2009}. For the numerous components described in Sec.~\ref{sec:components}, if users want to apply more components in their data videos and have greater control over their generation and coordination, they need to learn more tool-specific operations. To simplify this process, existing tools design new paradigms to optimize necessary operations, minimizing the user's learning curve and control requirements, while also leveraging AI to automate parts or the entirety of the process. For learnability, we analyzed the input and output of each work, focusing on how each work transforms diverse inputs into different outputs (data video components and their coordinations). As illustrated in Fig.~\ref{fig:learnability}, the transformation process can be categorized into four modes, which vary in the degree of human and AI involvement:

\review{Q11}\revise{As shown in Fig.~\ref{fig:learnability}, existing tools accept various materials as input and transform them into coordinated data video components. 
For each component and coordination in Sec.~\ref{sec:components}, early tools focused on optimizing manual interactions to reduce human effort, giving users absolute control over the process. 
The advancement of AI further automates some tasks, but introduces new challenges: users need to interact not only with the creative product but also with AI to guide video creation. The complexity introduced by different parties drives research into new design paradigms.
Thus, for creating and coordinating each component, these tools assign different roles to humans and AI. Understanding the interplay between human and AI roles is crucial to advance creation tools and design paradigms.
} 
By analyzing each tool‘s input and output, we identify four modes for transforming diverse inputs into coordinated data video components:

% \vspace{-3px}
\begin{itemize}
\item 
\textit{Original Mode}: No transformation occurs; inputs are directly translated to outputs without modification.
\item 
\textit{Human-Led Mode}: Users manually control the transformation, making precise adjustments through interfaces to achieve the desired results.
\item 
\textit{Mixed-Initiative Mode}: A combination of human and AI efforts where the user provides inputs and guidance, while AI aids in automating certain aspects of the transformation.
\item 
\textit{AI-Led Mode}: AI autonomously handles the transformation, without the need for user intervention.
% and allowing for efficient generation of components and their coordination.
\end{itemize}
% \vspace{-3px}

Data video creation is a complex process. Over the years, these tools have continuously evolved, with various design paradigms emerging across different transformation modes (Sec.~\ref{sec:learnability}) to facilitate data video authoring. Understanding and reflecting on these paradigms for each component's creation and coordination (Sec.~\ref{sec:expressivity}) is crucial for helping new researchers grasp the topic and inspire more effective future designs and developments.

\begin{table*}[t]
% \small
\centering
\caption{Interaction Paradigms for visualization creation in data video tools (Sec.~\ref{sec:vis}).}
% \vspace{-10px}
\label{tab: visualization}
\renewcommand\arraystretch{0.8}
\setlength{\tabcolsep}{1mm}{
\begin{tabular}{lll}
% \rowcolor[HTML]{EFEFEF} 
\toprule
\textbf{Mode} & \textbf{Paradigm} & \textbf{Representative Tools} \\
\midrule
Original & Upload &  
\cite{ 
Shen2024b,
dataplaywright,
Kim,
Kim2020,
Li2021c,
Lu2020a,
Tseng2024,
Yu2010,
Wang2019a,
Ge2020,
Lai2020a,
Wang2021d,
Lee,
Thompson2021,
Li2023b,
Zong2022,
Nam2024,
Conlen2022,
Latif2022,
Hall2022,
Zhao2022,
dataplayer,
Ying2023,
Femi-Gege2024,
Shi2023b,
wonderflow}\\
% \arrayrulecolor{tablerowcolor} \cmidrule(lr){1-3}
\arrayrulecolor{tablerowcolor}\midrule
 & Programming &  
\cite{Bostock2011}  \\ 
\arrayrulecolor{tablerowcolor} \cmidrule{2-3}
\multirow{-2.5}{*}{Human-Led} & Widgets &  
\cite{Yao2024, Tong2024} \\ 
% \arrayrulecolor{tablerowcolor} \cmidrule(lr){1-3}
\arrayrulecolor{tablerowcolor}\midrule
& Natural Language & 
\cite{Zhou2024,Chen2022c,DataParticles2023} \\
\arrayrulecolor{tablerowcolor} \cmidrule{2-3}
& Sketch & 
\cite{Lee2013} \\
\arrayrulecolor{tablerowcolor} \cmidrule{2-3}
& Example & 
\cite{Amini2017} \\
\arrayrulecolor{tablerowcolor} \cmidrule{2-3}
\multirow{-5}{*}{Mixed-Initiative} & Widgets & 
\cite{Tang2022, Lu2020b} \\
% \arrayrulecolor{tablerowcolor} \cmidrule(lr){1-3}
\arrayrulecolor{tablerowcolor}\midrule
& Heuristic Rules & 
\cite{
Wang2016e,
Li2020c,
Chen2022h,
Lin2023b,
Chen2023d,
Shin2022} \\
\arrayrulecolor{tablerowcolor} \cmidrule{2-3}
\multirow{-2.5}{*}{AI-Led} & Optimization & 
\cite{Shi2021a, NarrativePlayer} \\
% \arrayrulecolor{tablerowcolor2}\cmidrule(lr){1-3}
\bottomrule
\end{tabular}}
  % \vspace{-12px}
\end{table*}

\section{Analysis of Existing Tools}\label{sec:tools}
This section will provide an in-depth analysis of existing tools within the above framework.
% from the perspectives of expressivity and learnability. 
Tab. \ref{tab:tools} presents an overview of these tools, categorized into three groups as described in Sec.~\ref{sec:expressivity}. 
Each tool is labeled with its transformation mode (Sec.~\ref{sec:learnability}) for the components and coordination relationships outlined in Sec.~\ref{sec:expressivity}. 
The lower part of Tab. \ref{tab:tools} provides statistical information.
% for each transformation mode.
In the following subsections, for each component and coordination relationship, we will first analyze and summarize the key design paradigms within each transformation mode, and then present our reflections.
% We note that the section will only focus on components (Sec.~\ref{sec:components}) and coordination relationships (Sec.~\ref{sec:expressivity}) closely related to data video production.
% As expressivity increases, these tools strive to lower learnability by implementing intuitive interfaces and automation features. For each type of data video component and coordination relationship, we explore the roles of humans and AI and further summarize specific paradigms in the transformation process from user input to output. In each subsection, we provide a summary of our findings.

\subsection{Animation Unit}\label{sec:animationUnit}
Animation unit is the smallest dynamic representation of visuals. It involves creating both the visual elements and motion effects, as well as their semantic alignment.

\subsubsection{\visual{Visualization}}\label{sec:vis}
% Most existing data video tools (24/45) require users to upload visualizations or prepare them beforehand, with limited options for modifying these original visualizations. 
As shown in Tab.~\ref{tab:tools} and Tab.~\ref{tab: visualization}, most existing data video tools (26/46) require users to \textbf{upload or prepare visualizations in advance}, offering limited options for modification. 
% This reduces tool complexity and the learning curve to some extent. 
This avoids the visualization authoring process, thereby reducing tool complexity to some extent.
However, the upload formats vary across different tools and require different handling. 
For instance, 
% SVG is a common input format that organizes hierarchical visualization information. 
Data Animator~\cite{Thompson2021} and Narvis~\cite{Wang2019a} accept standard SVG files that organize hierarchical visualization information, while Canis~\cite{Ge2020} and Cast~\cite{Lee} utilize a data-enriched variant of SVG. Furthermore, WonderFlow~\cite{wonderflow} and Data Player~\cite{dataplayer} enhance SVGs by incorporating visualization structure information. Similarly, InfoMotion~\cite{Wang2021d} leverages graphics within slideshows.
In addition, Gemini~\cite{Kim, Kim2020} and Animated Vega-Lite~\cite{Zong2022} accept Vega-Lite specifications~\cite{Satyanarayan2017}, which is a declarative visualization programming language. 
% They also integrate a rendering engine within the system. 
GeoCamera~\cite{Li2023b} and V-Mail~\cite{Nam2024} focus on 3D visualizations, requiring a camera to present different viewpoints. 
Vis-Annotator~\cite{Lai2020a} analyzes uploaded raster images and extracts features using computer vision techniques. Additionally, Shi \etal~\cite{Shi2023b} directly handle existing data videos that contain visualizations.

Furthermore, 20 out of 46 tools create visualizations from uploaded datasets and related information. 
Notably, \textbf{manual tools} like D3~\cite{Bostock2011} require complex programming, allowing for fine-grained customization. To lower the learning curve, SwimFlow~\cite{Yao2024} and VisTellAR~\cite{Tong2024} encapsulate necessary low-level functions into tool-specific widget operations for intuitiveness, enabling users to configure their visualizations interactively.

With AI advancements, some tools facilitates \textbf{mixed-initiative visualization authoring}. These tools enable users to communicate their intents through various means, including messages~\cite{Zhou2024, Chen2022c, DataParticles2023}, sketches~\cite{Lee2013}, examples~\cite{Amini2017}, and widgets~\cite{Tang2022, Lu2020b}. 
For instance, Epigraphics~\cite{Zhou2024} allows users to input narrative messages and automatically generates asset recommendations for selection and manipulation in animated storytelling. 
SketchStory~\cite{Lee2013} interprets users' sketch gestures to initiate charts and automatically completes them by combining user visual examples with underlying data.
Furthermore, data2video~\cite{Lu2020b} and Yu \etal~\cite{Yu2010} also support data exploration during the authoring process but specifically focus on time-series data.

To unleash humans in visualization creation during video authoring, some tools provide \textbf{fully automatic visualization generation abilities} upon dataset loading. 
These tools typically rely on fixed parsing patterns that focus on specific visualization types and narratives, 
% tailored to particular narratives, 
such as dynamic scatter plots Rosling
% \footnote{The best stats you've ever seen | Hans Rosling. \url{https://www.youtube.com/watch?v=hVimVzgtD6w}} 
presented for showcasing changes in time-series events~\cite{Shin2022}, animated bar charts for ranking data~\cite{Li2020c}, and icon arrays or jittered scatter plots for illustrating data transformations~\cite{Pu2021}.
Furthermore, tools that enhance real-world scene videos with visualizations often have defined design spaces~\cite{Chen2022h, Lin2023b, Chen2023d}. Utilizing computer vision techniques, these tools process videos to extract features and pre-generate corresponding visualizations or data-driven pictograph candidates within the design space.
In addition to heuristic rules, AutoClips~\cite{Shi2021a} and Narrative Playter~\cite{NarrativePlayer} employ an optimization method to map provided or extracted data facts to specific animated visualizations, minimizing the overall transition cost for the visualization sequence. 

\textbf{\textit{Reflection}:}
User-uploaded visualizations are prevalent in data video creation, as visualization design itself is a significant research topic independent of video production.
\textbf{Different input formats impose varying requirements on users when preparing visualizations.} For example, images are the most common in everyday life, while SVGs require some design expertise, programming necessitates coding knowledge, and 3D visualizations demand advanced design and development skills. 
% Although current tools can accommodate a range of inputs, achieving more expressive data videos still relies on the quality of user-generated visualizations. 
Although current tools support various inputs, the expressivity of data videos largely depends on the quality of user-provided visuals. 
% ~\review{Q3}\revise{Moving forward, it is crucial to help end-users efficiently prepare visualization materials while also enhancing tools' capabilities to handle diverse input formats effectively.}

When creating visualizations within data video tools, \textbf{there is generally a trade-off between the automation level and visualization expressivity. }
This arises because different tools are designed for various scenarios. For example, many human-led and mixed-initiative tools primarily serve as authoring aids, helping users effectively translate their intents into visualizations~\cite{Tong2024, Thompson2021, Lee, Amini2017}. 
Full-automatic tools often play a dual role in data exploration, enabling users to visualize data rapidly~\cite{Latif2022, Shin2022, Shi2021a}. 
As a result, the visualizations generated by automated tools tend to be basic charts to illustrate data insights, while those manually fine-tuned or co-created with AI are generally more aesthetically pleasing and aligned with human intents.
% As a result, the visualizations generated by automated tools tend to be simpler compared to those that are manually fine-tuned or uploaded.
% and immersive visualization presentation

~\review{Q3}\revise{Enabling end-users to upload visualizations opens up design options but requires higher system parsing capabilities. 
Creating visualizations within data video tools confines users to their capabilities but can improve system discoverability and bridge the gap between visualization authoring and video creation.}
An important vision for data video tools is to help users efficiently prepare desired visual materials while enhancing tools' capacity to effectively manage diverse input formats. 
Prior to this, balancing the two objectives tailored to scenario contexts is essential.

\review{Q2}\revise{
\textbf{\textit{Summary:}}
In visualization creation (Tab.~\ref{tab: visualization}), user-uploaded visualizations dominate data video creation tools. Users can also employ various paradigms to convey intents; typically, ambiguous intents require more AI intervention for interpretation, while specific intents lead to relatively more expressive visualizations. 
% user-uploaded visualizations account for the majority in data video creation tools. Balancing automation and user input is key in creating visually compelling data videos, where different tools cater to varied user needs and levels of expressivity.
}

\begin{figure*}[t]
  \centering
    \includegraphics[width=\linewidth]{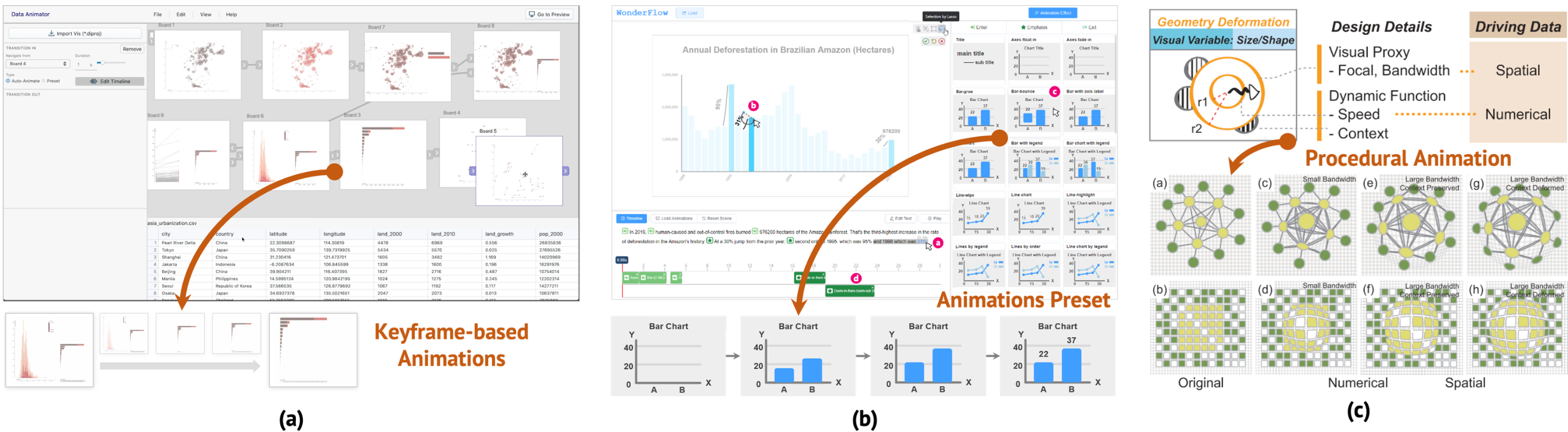}
% \vspace{-10px}
    \caption{Animation authoring paradigms: (a) keyframe animations~\cite{Thompson2021}; (b) animation preset~\cite{wonderflow}; and (c) procedural animations~\cite{Lu2020a}.}
\label{fig:animation}
  % \vspace{-10px}
\end{figure*}

\begin{table*}[t]
% \small
\centering
\caption{Paradigms for animation creation in data video tools (Sec.~\ref{sec:animation}) and different roles of humans and AI for application.}
\label{tab: animation}
% \vspace{-10px}

% \renewcommand\arraystretch{1}
\setlength{\tabcolsep}{1.5mm}{
\begin{tabular}{c l l l}
% \toprule
\hline
& \textbf{Keyframe Animation} & \textbf{Presets and Templates} & \textbf{Procedural Animation} \\\hline
% \midrule
Human-Led 
& \cite{ 
Bostock2011,
Kim2020,
Wang2019a,
Ge2020,
Zong2022,
Tong2024, Zhao2022}
& \cite{ 
wonderflow,
aftereffects, Li2020c}
& \cite{
Bostock2011}
\\\arrayrulecolor{tablerowcolor}\midrule

Mixed-Initiative 
& \cite{ 
Amini2017,
Lee,
Shen2024b,
Thompson2021,
Pu2021,
Li2021c,
Li2023b,
Nam2024}
& \cite{
dataplaywright,
Zhou2024,
Tseng2024,
Lee2013,
Amini2017,
Lu2020b,
DataParticles2023,
Li2023b,
Yao2024,
Conlen2022,
Chen2022c,
Andrews2024,
Lin2023b,
Chen2023d} 
& \cite{
Chen2022c,
Andrews2024,
Hall2022,
Femi-Gege2024,
Yao2024,
Lin2023b,
Chen2023d}
\\\arrayrulecolor{tablerowcolor}\midrule

AI-Led 
& \cite{ 
Kim,
Shi2021a}
& \cite{ 
Wang2016e,
dataplayer,
Shi2021a,
Shi2023b,
Ying2023,
Chen2022h,
Wang2016e,
Lai2020a,
NarrativePlayer,
Wang2021d,
Tang2022,
Shin2022,
flourish} 
& \cite{ 
Lu2020a,
Yu2010}
\\\arrayrulecolor{tablerowcolor2}\bottomrule
% \bottomrule
\end{tabular}}
  % \vspace{-10px}
\end{table*}

\subsubsection{\motion{Motion}}\label{sec:animation}
The animation landscape features three main authoring paradigms~\cite{Thompson2020}, as shown in Fig.~\ref{fig:animation}: 
\textit{Keyframe animation} allows precise control by defining specific frames at certain time points. Animations are generated by tweening between keyframes~\cite{Thompson2021, Lee} (Fig.~\ref{fig:animation}-a). 
\textit{Presets} simplify the animation process with predefined structures and effects, making animation accessible but potentially limiting creativity~\cite{Amini2017, Shi2021a} (Fig.~\ref{fig:animation}-b). 
\textit{Procedural animation} generates dynamic motions through defined behavior parameters, ideal for simulating complex movements~\cite{Lu2020a, Kazi2014} (Fig.~\ref{fig:animation}-c). 

% offering fluidity and adaptability without manual intervention, 
It is common for an animation tool to integrate one or more of the aforementioned authoring paradigms, and leverage diverse interaction modalities and AI capabilities to enhance the user experience. 
% Meanwhile, advancements in AI have further optimized and upgraded these applications, enabling more efficient and flexible animation creation.
Tab.~\ref{tab: animation} summarizes the application of these paradigms, highlighting the roles of both humans and AI.

Programming is a common \textbf{manual method for animation creation}.
The programming languages vary in design paradigms and syntax granularity, tailored for different applications.
% , serving as a bridge between interactive layers and system implementation. 
% Different syntax granularities allow professionals to customize animations effectively.
For example, D3~\cite{Bostock2011} employs an imperative approach with a functional programming style. This allows developers to define desired outcomes while the library manages updates and transitions between keyframes using \textit{transition} and \textit{ease} functions. D3 also excels in procedural animation, dynamically generating motions in response to data changes through its robust data-binding and SVG manipulation capabilities.
There are several keyframe-based declarative languages with unique features. For example, Canis~\cite{Ge2020} allows users to partition selected marks in an SVG visualization and animate the mark groups. Animated Vega-Lite~\cite{Zong2022} encodes time as an independent channel, mapping data fields to animation keyframes. Gemini~\cite{Kim2020} and AniVis~\cite{Li2021c} define transition ``steps'' for visual components across two provided visualizations.
Fidyll~\cite{Conlen2022} reduces the programming learning curve by enabling users to compose their data narratives in a high-level markup language, which is then parsed into a machine-readable JSON structure. 
Overall, programming languages enhance user-friendliness by evolving from imperative to declarative and markup styles. 
This evolution simplifies syntax while requiring enhanced parsing and inference capabilities in compilers.
Different languages serve as a bridge between interactive layers and system implementation, suitable for different application scenarios.
% This allows compilers to simplify syntax and tailor grammars to specific applications, serving as a bridge between user interaction and system implementation.
% This evolution strengthens compilers to simplify syntax while recognizing that different grammars are suited for various application scenarios.

In addition to programming with keyframe and procedural animations, selecting specific animation templates is another major manual animation authoring paradigm, as commonly seen in commercial software like Microsoft PowerPoint and Adobe After Effects~\cite{aftereffects}. 
Moreover, WonderFlow~\cite{wonderflow} includes a structure-aware animation library that lets users apply suitable animations to the selected visual elements (Fig.~\ref{fig:animation}-b). 
% there are two major manual animation paradigms. 
% The first involves manually selecting specific animation templates, as commonly seen in commercial software like Microsoft PowerPoint and Adobe After Effects~\cite{aftereffects}. For instance, WonderFlow~\cite{wonderflow} includes a structure-aware animation library that lets users apply suitable animations to the selected visual elements (Fig.~\ref{fig:animation}-b). 
In addition, some tools integrate pre-configured motion effects that require user interaction to trigger the appearance or disappearance of elements at specific times~\cite{Wang2019a, Tong2024, Zhao2022}.
% These effects are typically simple, involving basic transitions such as appearing or disappearing, and will be further explored in Section 3.3.

\textbf{The mixed-initiative approaches in animation unit creation}, similar to the human-computer collaboration methods in visualization creation, empower users to articulate their animation intents, which are then automatically parsed into animation implementations.
Among them, natural language is an important interaction modality~\cite{NLISurvey}, encompassing two primary modes. 
The first is command-based~\cite{Tseng2024, Andrews2024}, exemplified by tools like Keyframer~\cite{Tseng2024}, which allows users to input natural language commands (\eg ``\textit{move the yellow star from the bottom to the top}'') to generate SVG animations. Users can then iteratively refine these animations through conversational adjustments. The second mode is narration-based~\cite{Zhou2024, DataParticles2023, Chen2022c}. For instance, Sporthesia~\cite{Chen2022c} focuses on sports videos, users can input commentary (\eg ``\textit{Federer hits a backhand down the line}''), and the system will automatically detect visualizable entities within the narration and generate embedded visualizations within the video.
Data Playwright~\cite{dataplaywright} introduces a novel paradigm that allows users to express their narrative and authoring intents in a unified format called annotated narration.
Rather than inferring animation intents, GeoCamera~\cite{Li2023b} summarizes narrative intents related to camera use (\eg emphasize a target, overview multiple targets, make a comparison, \etc).
% , and explicitly presents them in the interface for selection.
They are presented on the interface for selection, and the system combines them with user-specified keyframes to generate camera movements automatically.
Similarly, keyframes are also widely used in other tools as narrative intents~\cite{Lee, Shen2024b, Thompson2021, Pu2021, Li2023b, Nam2024, Amini2017}, since they are the external manifestations of the story (which will be further discussed in Sec.~\ref{sec:vis-vis}). These tools require users to manually specify keyframes for specific visualizations or elements, after which the system generates transition animations automatically.
Furthermore, data insights or events, as core elements of data storytelling, are also a crucial intent type. For instance, Roslingifier~\cite{Shin2022} allows users to select interested data events, and then automatically generate or update animations. 
Additionally, another set of animations is driven by human gestures~\cite{Hall2022, Lee2013, Yao2024, Femi-Gege2024}. They focus on the appearance and disappearance of visual elements based on the relationship between gestures and visualizations.

\textbf{Automatic animation generation methods} typically integrate an animation library and utilize AI to suggest appropriate effects based on users' initial inputs, including raw data~\cite{Shin2022, Latif2022}, data facts~\cite{Shi2021a}, visualization~\cite{Lu2020a, Wang2021d, Ying2023}, visualization sequence~\cite{Kim}, visualization with text descriptions~\cite{dataplayer, Lai2020a}, and video~\cite{Shi2023b, Chen2022h}. 
These inputs correspond to various narrative roles of automatically generated animations. 
% and Roslingifier~\cite{Shin2022}
For instance, tools like Talking Realities~\cite{Latif2022} and Yu \etal~\cite{Yu2010} support data exploration and generate animated visualizations, and AutoClips~\cite{Shi2021a} maps data fact sequences to an animated clip library. The primary purpose of these animations is to visually present and explain data insights.
Lu \etal~\cite{Lu2020a} use data-driven animations to emphasize certain attributes of the static visualization. The model employs a procedural animation paradigm that encodes a set of messages into five design components with parameters (Fig.~\ref{fig:animation}-c).
% and Yu \etal~\cite{Yu2010}
Gemini 2~\cite{Kim} builds on its predecessor, Gemini~\cite{Kim2020}, by further automating the generation of transition animations. 
Vis-Annotator~\cite{Lai2020a} automates the annotation of visualizations based on textual descriptions, generating animations for step-by-step presentations of sentences. 
Similarly, Shi \etal~\cite{Shi2023b} automate the creation of video annotations based on the narration of existing animated data videos.
% , enhancing the readability of the videos. 
% These systems illustrate the diverse roles that animations can play in reinforcing data narratives and improving user comprehension.

\begin{table*}[t]
% \small
\centering
\caption{Paradigms for coordinating visualization and motion in data video tools (Sec.~\ref{sec:VisAni}).}
\label{tab: VisAni}
% \vspace{-10px}
\renewcommand\arraystretch{1}
\setlength{\tabcolsep}{1mm}{
\begin{tabular}{cm{12cm}}
% \rowcolor[HTML]{EFEFEF} 
\toprule
\textbf{Mode} & \textbf{Representative Tools} \\
\midrule
Human-Led & 
\cite{ 
Bostock2011,
Wang2019a,
Ge2020,
Li2020c,
Tong2024,
Conlen2022,
Zhao2022}\\
% \arrayrulecolor{tablerowcolor} \cmidrule(lr){1-2}
\arrayrulecolor{tablerowcolor}\midrule
Mixed-Initiative &
\cite{ 
Tseng2024,
Hall2022,
Femi-Gege2024}\\
% \arrayrulecolor{tablerowcolor} \cmidrule(lr){1-2}
\arrayrulecolor{tablerowcolor}\midrule
AI-Led &
\begin{tabular}[c]{@{}l@{}}
Visual Feature:
\cite{
dataplayer,
Ying2023,
Kim, 
Kim2020, 
Li2021c,
Lee2013,
Amini2017,
Wang2021d,
Thompson2021,
Wang2016e,
Tang2022,
dataplaywright,
NarrativePlayer,
wonderflow} 
\\
Semantic Feature:
\cite{
Lu2020a,
Zhou2024,
Yu2010,
Lu2020b,
Lai2020a,
Shi2021a,
Lee,
Shen2024b,
Pu2021,
Chen2022c,Chen2023d,Chen2022h,Lin2023b,Lin2023a,Yao2024,
DataParticles2023,
Li2023b,
Zong2022,
Nam2024,
Latif2022,
Shin2022,
Shi2023b,
Andrews2024}
\end{tabular}\\
\arrayrulecolor{tablerowcolor2}\bottomrule

\end{tabular}}
  % \vspace{-10px}
\end{table*}

\textbf{\textit{Reflection}:}
\textbf{The three main animation authoring paradigms offer distinct strengths and challenges for human and AI involvement.}
% , and they can be better integrated in the future.
Keyframe animation typically requires human involvement to program or provide keyframe information. Existing automation technologies primarily focus on detecting relationships between keyframes and recommending suitable transition effects, as well as converting other data into keyframes. 
While the flexibility of keyframes allows for greater user customization and expressivity, it also complicates the development of clear patterns for widespread automation.
% While the vast space of keyframes allows for greater user customization for better expressivity, it complicates the specification of clear patterns for widespread automation. 
In contrast, animation templates often feature more identifiable patterns, making AI assistance more feasible. These patterns can be flexibly integrated and extended, but current tools are limited by a narrow range of templates, restricting expressivity and highlighting the need for more versatile workflows. 
% In contrast, animation templates usually feature more identifiable patterns and can be flexibly integrated and extended, making AI assistance more feasible. However, current interactive authoring tools often rely on limited templates, constraining expressivity and highlighting the need for more versatile workflows.
Compared to the other two paradigms, procedural animations in data visualization have received comparatively little attention. 
% This represents a promising area for future exploration.
% mainly enhancing pre-recorded videos, 
Some combinations of these paradigms have already improved user experience, such as specifying keyframes through templates~\cite{Amini2017}. Future research could explore more effective combinations of these paradigms, as well as the development of new authoring paradigms.

% The animation process often reveals discernible patterns, leading to a superiority of AI assistance in data video tools. 
% , demanding substantial design expertise and engineering effort. 
% Overall, animation creation is one of the most complex tasks in data video production, but the animation process often follows recognizable patterns, making AI assistance a dominant approach. The effectiveness of AI in this context is largely influenced by the specificity of user input.
Animation is one of the most complex aspects of data video production, often requiring significant design expertise and technical effort. However, due to the repetitive and pattern-based nature of many animation tasks, \textbf{AI assistance in animation creation has become widely adopted but the effectiveness of AI in this context is largely influenced by the specificity of user input.}
% The animation process frequently exhibits recognizable patterns, making it well-suited for simplification through AI assistance.
\review{Q3}\revise{As user input becomes less specific (\eg providing only a dataset), the animation task becomes more open-ended, giving the AI greater interpretative flexibility but also increasing the risk of misalignment with user intent. Conversely, when users provide more detailed inputs (such as visualizations or text descriptions), clearer patterns emerge, allowing the AI to better align with the user's goals.} For example, if the user supplies only two keyframes, the animation will likely prioritize the transition between those frames.
Moving forward, future research should explore more sophisticated mechanisms for balancing AI-driven automation with user control. 
\revise{One important direction is to enhance the feedback mechanism in production. Current research primarily focuses on sequential processes, overlooking the loop between the creator, tool, and audience.}
% Current interactive authoring tools often rely on limited templates, constraining expressivity and underscoring the need for versatile pipelines. Procedural animations remain neglected in the data video domain, presenting a ripe area for future exploration. Additionally, the existing tools predominantly generate animations from visualizations, suggesting a potential for future research into animation-driven coordination.

\review{Q2}\revise{
\textbf{\textit{Summary:}}
In motion creation (Tab.~\ref{tab: animation}), three primary paradigms exist: Keyframe animation typically requires human input of visual keyframes, animation templates offer reusable patterns for AI support, and procedural animations involve parameter configuration, suitable for AI but relatively overlooked. 
Tools can integrate one or more paradigms.
}

\subsubsection{\visual{Visualization}---\motion{Motion}}\label{sec:VisAni}
Motion creation involves both motion effects and applied visual elements, highlighting the importance of their alignment.
As shown in Tab.~\ref{tab: VisAni}, \textbf{current approaches primarily automate visual or semantic alignment in animation production}, in addition to the programmed animations and manual assignment of simple appearance and disappearance effects ~\cite{Wang2019a, Ge2020, Li2020c, Tong2024, Conlen2022, Zhao2022}, as well as mixed-initiative gesture-driven motion discussed previously~\cite{Tseng2024, Hall2022, Femi-Gege2024}. 
These approaches typically conduct formative studies to establish tailored design spaces and develop specialized animation libraries or presets.
% They typically involve conducting a series of formative studies to establish a tailored design space and develop a specialized animation library.
% current approaches to achieving this alignment typically involve conducting a series of formative studies to identify a tailored design space and develop a specialized animation library. 
These efforts focus on various aspects, such as visualization structure~\cite{wonderflow,dataplayer}, geographic camera movements~\cite{Li2023b,Nam2024}, visualization enhancement~\cite{Lu2020a}, visualization presentation~\cite{Wang2021d,Wang2019a}, graphical annotations~\cite{Shi2023b}, sports video augmentation~\cite{Chen2022c,Chen2023d,Chen2022h,Lin2023b,Lin2023a}, and embedding within real-world scenes~\cite{Tang2022,Tong2024}.
% Animation libraries often include a range of presets. 
For example, DataClips~\cite{Amini2017} features a library of clips combining high-level visualization types with animation types, created from an analysis of existing data videos. WonderFlow~\cite{wonderflow} and Data Player~\cite{dataplayer} have developed visualization structure-aware animation libraries,
% to align visual elements with their shapes and semantics, 
offering presets like ``grow'' for bar, ``wheel'' for pie, and ``wipe'' for line. They also support combinations of component-level animations, such as a pie chart's ``wheel'' effect occurring simultaneously with a legend ``fly-in'', and axes appearing first, followed closely by the marks. 
In addition, WonderFlow~\cite{wonderflow} includes filtering logic that ensures users can only choose appropriate animations for the selected visual elements. Data Player~\cite{dataplayer} gathers high-level animation design constraints through interviews to inform its animation recommendation module.
GeoCamera~\cite{Li2023b} analyzes existing geographic data videos to compose a design space for camera movements based on geospatial targets and narrative intents. Lu \etal~\cite{Lu2020a} derive a design space for data-driven animated effects through an empirical user study, examining the impact of and user preferences for three versatile effects to enhance static visualizations. Sporthesia~\cite{Chen2022c} analyzes 155 sports videos and their commentaries to determine which text entities can be visualized and animated for augmented sports video creation.

\textbf{\textit{Reflection}:}
\textbf{Many tools have integrated application-specific frameworks or design spaces for visualization and motion alignment}, establishing clear patterns that AI can learn to offer system recommendations. 
\review{Q3}\revise{This enables end-users to focus on their creative intent without concerning themselves with specific alignment details.}
Additionally, current tools primarily generate animations from visualizations, indicating potential for future research into animation-driven coordination. For instance, users could select preferred animation templates or provide references, and the system then automatically chooses corresponding visual elements. 

% While existing design spaces or animation libraries are specific to particular tools, 
In fact, many empirical studies have more generally explored the coordination between visualization and animation, covering various areas, including affective animation design~\cite{Lan2023, Lan2022}, cinematic effects~\cite{Xu2023b, Xu2022, Conlen2023}, visual narrative~\cite{Shi2021b}, animation transitions~\cite{Rodrigues2024, Dragicevic2011, Robertson2008, Tang2020}, visual cue preferences~\cite{Kong2019, Kong2017a}, and 3D data videos~\cite{Yang2023a}.
However, \textbf{few existing empirical research results have been directly applied to tool development.} A significant reason for this is that many existing design spaces or guidelines are high-level descriptions rather than computable solutions, making it challenging to address specific domain issues comprehensively. Therefore, a key future direction is bridging the gap between empirical research and technical implementation.

\review{Q2}\revise{
\textbf{\textit{Summary:}}
In the coordination of visualizations and motions (Tab.~\ref{tab: VisAni}), existing tools primarily automate the visual or semantic alignment of these two aspects within predefined or application-specific design spaces.
}

\begin{table*}[t]
\small
\centering
\caption{Narrative sources in data video tools (Sec.~\ref{sec:narrative}).}
\label{tab: narrative}
% \vspace{-10px}

\renewcommand\arraystretch{0.8}
\setlength{\tabcolsep}{3mm}{
\begin{tabular}{lll}
% \rowcolor[HTML]{EFEFEF} 
\toprule
\textbf{Mode} & \textbf{Narrative Source} & \textbf{Representative Tools} \\
\midrule

& Text &  \cite{wonderflow, dataplayer, Lai2020a, DataParticles2023, NarrativePlayer, dataplaywright} \\
\arrayrulecolor{tablerowcolor} \cmidrule{2-3}
\multirow{-2.5}{*}{Original} 
& Video & \cite{Chen2022c, Chen2023d, Chen2022h, Lin2023b, Shi2023b, Yao2024, Andrews2024,Tang2022}\\
% \arrayrulecolor{tablerowcolor} \cmidrule(lr){1-3}
\arrayrulecolor{tablerowcolor}\midrule

& Sketch &  \cite{Lee2013, Femi-Gege2024, Hall2022, Zhao2022, Tong2024}  \\ 
\arrayrulecolor{tablerowcolor} \cmidrule{2-3}
\multirow{-2}{*}{Human-Crafted} 
& Keyframe Specification & \cite{
Amini2017,
Wang2019a,
Ge2020,
Lee,
Shen2024b,
Thompson2021,
Li2023b,
Zong2022,
Nam2024,
Conlen2022} \\
% \arrayrulecolor{tablerowcolor} \cmidrule{2-3}
% & Recording &  \cite{Femi-Gege2024, Hall2022, Zhao2022, Tong2024}  \\ 
% \arrayrulecolor{tablerowcolor} \cmidrule(lr){1-3}
\arrayrulecolor{tablerowcolor}\midrule

& Data & \cite{Latif2022,Shin2022,Yu2010,Wang2016e,Lu2020b,Li2020c} \\
\arrayrulecolor{tablerowcolor} \cmidrule{2-3}
& Visualization & \cite{Ying2023,Wang2021d} \\
\arrayrulecolor{tablerowcolor}\cmidrule{2-3}
& Data Query &  \cite{Pu2021}  \\ 
\arrayrulecolor{tablerowcolor} \cmidrule{2-3}
% \multirow{-2}{*}{Mixed-Initiative Transformed} 
\multirow{-5}{*}{AI-Generated} 
& Data Facts & \cite{Shi2021a} \\
\arrayrulecolor{tablerowcolor2}\bottomrule
\end{tabular}}
  % \vspace{-10px}
\end{table*}

\subsection{Animated Narrative}
The previous section introduced the construction practice of the minimal animation unit.
However, to convey a complete narrative, these units must be organized into a sequence. 
To achieve this, in addition to animation units, it is also essential to coordinate between visualizations and between real-world scenes and visualizations.
% , focusing primarily on the presentation of visual elements with dynamic effects. 

\subsubsection{\narrative{Narrative}}\label{sec:narrative}

In data videos, a narrative can be described as an organized sequence of events that shape visual components and their timing~\cite{Riche2018}. As discussed in Sec.~\ref{sec:visualcomponents}, 
% visual components are the foundation for conveying a narrative. 
generally, visualizations are used to convey data narratives, while real-world scenes and pictographs provide contextual information. 
When composing narratives, it is important to consider the information each visual component conveys, along with their sequence and structure~\cite{Li2023c}.
\textbf{In data video tools, narratives differ significantly in their sources and specificity}, requiring tool adaptations to enhance or complete the narrative. 
As shown in Tab.~\ref{tab: narrative}, some tools allow users to upload a complete narrative, in the form of narration text~\cite{wonderflow, dataplayer, Lai2020a, DataParticles2023} or video~\cite{Chen2022c, Chen2023d, Chen2022h, Lin2023b, Shi2023b, Yao2024}, and the tool further enhances the original narrative with visuals. 
% Humans can also craft their narratives through sketch and keyframe specifications
For human-crafted narratives, one category involves character recording, where users compose the narrative through real-time sketch-based presentations~\cite{Femi-Gege2024, Hall2022, Zhao2022}. 
Another important set of work allows users to specify each keyframe clearly to construct the narrative, followed by completing the transitions between keyframes~\cite{Amini2017, Wang2019a, Ge2020, Lee, Thompson2021, Li2023b, Zong2022, Nam2024, Conlen2022}. 
% These will be discussed in Sec.~\ref{sec:real-vis}. 
Furthermore, tools with AI-generated narrative capabilities allow users to input intermediate results from raw data, such as sequences of data facts~\cite{Shi2021a} and data queries~\cite{Pu2021}, which are then transformed into data videos. Additionally, there are cases where the narrative is not predefined and needs to be automatically extracted from the data~\cite{Latif2022, Shin2022, Lu2020b} or from visualizations to be deconstructed~\cite{Ying2023, Wang2021d}.
% These will be discussed in Sec.~\ref{sec:vis-vis}.

\textbf{\textit{Reflection}:}
Narrative is the soul of data videos, conveying the essential message to the audience. 
\review{Q3}\revise{From a creative perspective, \textbf{most existing systems still place control of narratives in the hands of humans, with AI playing a supportive role} by generating narratives based on user-provided information, especially in scenarios where users have not yet formulated a clear narrative (\eg when only raw data is available).}
In addition, current tools support inputting various intermediate results, such as data facts~\cite{Shi2021a}, queries~\cite{Pu2021}, and visualizations~\cite{Ying2023}, to assist in narrative creation. 
\review{Q3}\revise{They help bridge the gap between the data analysis process and the storytelling process for end-users.} 
Future directions could further explore integrating the various stages of the data exploration~\cite{Biswas2022} and information visualization pipelines~\cite{Ltifi2020} with the storytelling workflow~\cite{Lee2015}, facilitating narrative crafting from data.

\review{Q2}\revise{
\textbf{\textit{Summary:}}
In narrative creation (Tab.~\ref{tab: narrative}), narrative information comes from various sources (user inputs) and differs in specificity. 
Most tools maintain human control over narrative crafting, while AI plays a supportive role.
}

\subsubsection{\visual{Visualization}---\visual{Visualization}}\label{sec:vis-vis}
% Much of the conveyed information in real-world scenes pertains to context, while 
Data narratives in data videos are primarily communicated through visualizations. The temporal pacing between keyframes in visualizations, driven by data, significantly impacts the narrative's effectiveness~\cite{Hullman2013a,remex}. This pacing typically manifests in two forms: the sequential linking of different visualizations, where each visualization encodes distinct data entities~\cite{Amini2017, Thompson2021}, and the presentation of various elements within a single visualization, such as through staggered appearances, emphasis, or visual cues~\cite{dataplayer, wonderflow, Lee}. 
As shown in Tab.~\ref{tab: VisVis}, the coordination of visualizations is closely tied to the user input discussed in Sec.~\ref{sec:vis}. If users provide input for the visualizations, most work involves manually ordering the visual narratives. Without user-provided visualizations, the sequence must be generated from other sources. 

% template selection~\cite{Amini2017},
Similar to animation practices, \textbf{manually composing or specifying keyframes} is a common paradigm, achieved through various interaction modalities like programming~\cite{Ge2020, Conlen2022, Zong2022}, manual ordering~\cite{Thompson2021}, or interactive selection~\cite{Wang2019a, Lee, Li2023b, Nam2024, wonderflow, Hall2022, Femi-Gege2024}. 
In addition, if users have uploaded video narratives, the challenge of coordinating visualizations shifts to determining their placement on the narrative timeline, which will be discussed in Sec.~\ref{sec:real-vis}.

Furthermore, some \textbf{mixed-initiative approaches} manually order other information, which is then mapped into keyframes~\cite{Pu2021, Amini2017, Lee2013}. For instance, Datamations~\cite{Pu2021} maps the results of each step in user-written data transformation code to plots or tables, and transitions between these data operations are animated accordingly. 
Furthermore, DataParticles~\cite{DataParticles2023} employs a block-based editing paradigm for authoring animated unit visualizations, where each block represents a visualization (\ie story piece), allowing for frequent iterative modifications by the user.

\textbf{AI-led visualization coordination }primarily involves three paradigms: data-driven generation~\cite{Shin2022, Yu2010, Wang2016e, NarrativePlayer}, visualization-based interpretation~\cite{dataplayer, Ying2023}, and text-oriented linking~\cite{Lai2020a, dataplayer}. 
First, a significant category is data-driven visualization generation, where extracted data insights serve as the backbone of the narrative and are then converted into a visualization sequence. For example, Roslingifier~\cite{Shin2022} automatically extracts important events from temporally changing data and displays them on a timeline for users to explore and organize.
Second, visualizations contain numerous elements that can be grouped based on visual or semantic features, forming interconnected narratives with temporal pacing. 
% An important area of work involves interpreting visualizations to create these interconnected narratives.
For instance, InfoMotion~\cite{Ying2023} recognizes the underlying information structures (\eg similar elements, repeating units, connectors, and semantic tags) in static data infographics and applies appropriate animation sequences to the established element groups.
Third, text-visual linking involves semantically analyzing narrative text and aligning related text segments with visual elements. For example, Vis-Annotator~\cite{Lai2020a} automatically extracts text segments to serve as visualization annotations and applies animations to enhance the presentation with narrative pacing.

\begin{table*}[t]
% \small
\centering
\caption{Paradigms for coordinating visualizations in data video tools (Sec.~\ref{sec:vis-vis}).}
\label{tab: VisVis}
% \vspace{-10px}

\renewcommand\arraystretch{0.8}
\setlength{\tabcolsep}{3mm}{
\begin{tabular}{lll}
% \rowcolor[HTML]{EFEFEF} 
\toprule
\textbf{Mode} & \textbf{Paradigm} & \textbf{Representative Tools} \\
\midrule

& Programming &  \cite{Ge2020,Conlen2022,Zong2022}  \\ 
\arrayrulecolor{tablerowcolor} \cmidrule{2-3}
% & Template Selection & \cite{Amini2017} \\
% \arrayrulecolor{tablerowcolor} \cmidrule{2-3}
& Interactive Selection & \cite{Wang2019a,Lee,Li2023b,Nam2024,wonderflow,Hall2022,Femi-Gege2024, Shen2024b} \\
\arrayrulecolor{tablerowcolor} \cmidrule{2-3}
& Manual Ordering &  \cite{Thompson2021}  \\ 
\arrayrulecolor{tablerowcolor} \cmidrule{2-3}
\multirow{-4.5}{*}{Human-Led} 
& Timeline Insertion & \cite{Chen2022h,Lin2023b,Chen2023d,Yao2024,Tong2024,Chen2022c,Andrews2024,Tang2022,Zhao2022} \\
% \arrayrulecolor{tablerowcolor} \cmidrule(lr){1-3}
\arrayrulecolor{tablerowcolor} \midrule

& Keyframe Mapping & \cite{Pu2021, Amini2017,Lee2013} \\
\arrayrulecolor{tablerowcolor} \cmidrule{2-3}
\multirow{-2}{*}{Mixed-Initiative} 
& Block-Based Editing & \cite{DataParticles2023} \\
% \arrayrulecolor{tablerowcolor} \cmidrule(lr){1-3}
\arrayrulecolor{tablerowcolor} \midrule

& Data-Driven Generation & \cite{Shin2022, Yu2010, Wang2016e, Shi2021a, Li2020c,Lu2020b,Latif2022,NarrativePlayer} \\
\arrayrulecolor{tablerowcolor} \cmidrule{2-3}
& Visualization-Based Interpretation & \cite{Wang2021d, Ying2023} \\
\arrayrulecolor{tablerowcolor} \cmidrule{2-3}
\multirow{-3.5}{*}{AI-Led} 
& Text-Oriented Linking & \cite{Lai2020a, dataplayer,Shi2023b, dataplaywright} \\
\arrayrulecolor{tablerowcolor2}\bottomrule
\end{tabular}}
  % \vspace{-10px}
\end{table*}

\textbf{\textit{Reflection}:}
% The relationships between visualizations are crucial for conveying a narrative. 
Similar to the reflections in Sec.~\ref{sec:narrative}, \textbf{for visualizations, manual coordination remains the dominant approach}, particularly through interactive selection of visual elements for each keyframe and manual insertion of visualization keyframes along the timeline. AI assistance can be categorized into two main approaches: if no visualizations are provided, AI generates coordinated visualizations based on other information; if visualizations are given, AI’s primary task is mapping visual keyframes onto the timeline, which can take various forms such as a chronological timeline or narration text~\cite{dataplayer, Ying2023}. 
Although various coordination paradigms have been investigated, an unexplored paradigm is the generation of new visualizations based on previous ones, guided by narrative or data insights. 

In addition, \textbf{current tools tend to prioritize technical authoring while offering limited creative support for data narrative crafting}. 
Such creative support can consider multiple dimensions, such as target audience, narrative structure, presentation style, etc. 
For instance, tools could provide suggestions on narrative pacing, tone, or visual styles depending on the intended audience. 
\review{Q3}\revise{Moreover, different types of users may require tailored creative support: novices may benefit from more structured guidance and templates, while experts might need flexible frameworks that allow for greater customization. Additionally, collaborative projects demand different types of support compared to individual efforts, as they require coordination of creative decisions across multiple contributors. }
% Expanding creative support in this direction could enhance the storytelling potential of data videos.

\review{Q2}\revise{
\textbf{\textit{Summary:}}
In the coordination of visualizations (Tab.~\ref{tab: VisVis}), related to visualization creation (Tab.~\ref{tab: visualization}), user-uploaded visualizations often require manual keyframe sequencing for narrative construction. In their absence, AI typically maps visualization sequences from alternative narrative sequence data.
% if users upload visualizations, the mainstream paradigm is manually sequencing keyframes to shape visual narratives. Without user-provided visualizations, the sequence is typically mapped from alternative narrative sequences by AI.
}

\begin{table*}[t]
% \small
\centering
\caption{Paradigms for coordinating real-world scenes and visualizations in data video tools (Sec.~\ref{sec:real-vis}).}
\label{tab: real-vis}
% \vspace{-10px}

\renewcommand\arraystretch{1}
\setlength{\tabcolsep}{3mm}{
\begin{tabular}{ccc}
\toprule
% \hline
When to Configure Visualization & Human & Mixed-Initiative \\
\midrule
% \hline
After Recording &  
\cite{Yao2024}&  
\cite{ 
Chen2022h,
Lin2023b,
Chen2023d,
Tang2022,
Chen2022c,
Andrews2024}
\\
% \hline
% \arrayrulecolor{tablerowcolor} \cmidrule(lr){1-3}
\arrayrulecolor{tablerowcolor} \midrule

During Recording &  
\cite{Hall2022, Femi-Gege2024, Zhao2022}&  
\\
% \hline
% \arrayrulecolor{tablerowcolor} \cmidrule(lr){1-3}
\arrayrulecolor{tablerowcolor} \midrule

Before Recording &  
\cite{Tong2024}& 
\\
\arrayrulecolor{tablerowcolor2}\bottomrule
% \hline
\end{tabular}}
  % \vspace{-10px}
\end{table*}

\begin{figure*}[t]
  \centering
    \includegraphics[width=\linewidth]{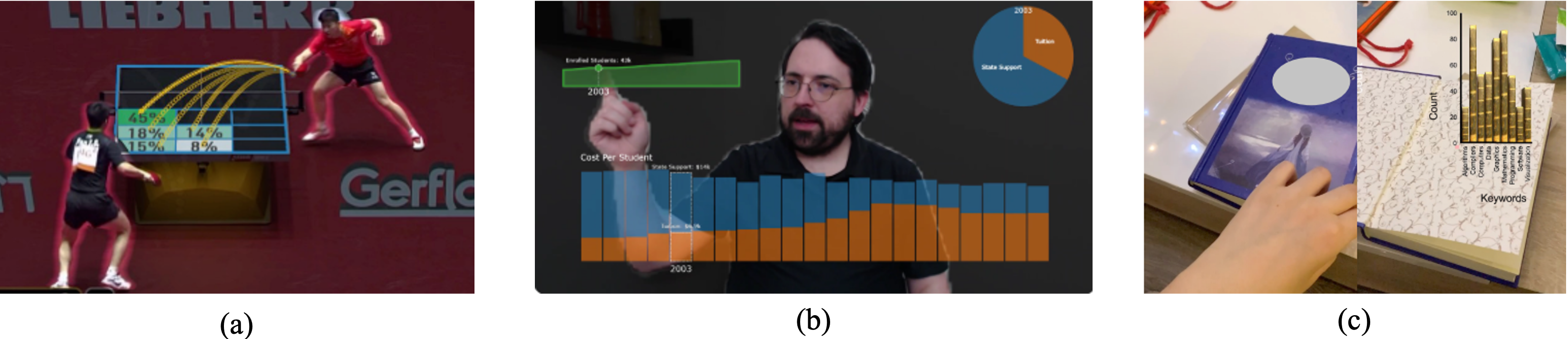}
% \vspace{-10px}
    
    \caption{Coordination between real-world scenes and visualizations: (a) Embedding visualizations into pre-recorded videos through post-editing~\cite{Chen2022h}; (b) Real-time recording to facilitate remote visualization presentations~\cite{Hall2022}; (c) Recording video and timing the insertion of pre-configured visualizations at key moments~\cite{Tong2024}.}
\label{fig:real}
  % \vspace{-10px}
\end{figure*}

\subsubsection{\visual{Real-World Scene}---\visual{Visualization}}\label{sec:real-vis}
Real-world scenes are essential for providing context in data videos, typically incorporated through general videos featuring such scenes.
\textbf{Visualizations can be configured after, during, or before video recording}, as shown in Tab.~\ref{tab: real-vis}.
% , either pre-recorded or captured live.
\textit{Pre-recorded general videos} come with fixed timelines and narratives, data video tools need to enhance these narratives by embedding appropriate visualizations at the right moments~\cite{Yao2022, Lin2023a}. 
Some approaches utilize computer vision techniques to automatically extract video features and generate suitable visual elements with video effects, particularly in sports videos (Fig.~\ref{fig:real}-a). 
These approaches are based on their respective design spaces but differ in application scenarios and interaction designs to trigger visual elements, thereby avoiding visual clutter and occlusion.
For example, VisCommentator~\cite{Chen2022h} and SwimFlow~\cite{Yao2024} integrate direct manipulation interfaces, allowing users to visualize sports data for specific entities at certain timestamps. Sporthesia~\cite{Chen2022c} further streamlines this process using natural language interfaces. AiCommentator~\cite{Andrews2024} supports conversational interactions, and iBall~\cite{Chen2023d} incorporates eye-gaze interactions to trigger specific visuals. Omnioculars~\cite{Lin2023b} addresses in-game dynamic data by embedding context-driven visualizations in live basketball games.

Beyond post-editing techniques, another important category of work focuses on \textit{real-time recording}, particularly in remote presentations involving characters, where the character's presence enhances the presentation's authenticity~\cite{Hall2022, Femi-Gege2024, Zhao2022}. In these scenarios, the narrative is driven by the presenter's gestures and speech while interacting with static charts (Fig.~\ref{fig:real}-b). For example, Hall \etal~\cite{Hall2022} overlay dynamic charts on a presenter’s webcam feed, utilizing continuous hand tracking to manipulate and emphasize chart elements, thus making the presentation more expressive and engaging.
Additionally, VisTellAR~\cite{Tong2024} introduces a novel pipeline that allows users to create and configure data visualizations \textit{before video recording}. Users can record videos with their smartphones, indicating through speech and gestures when visualizations should appear, and subsequently post-edit the visualizations after recording (Fig.~\ref{fig:real}-c).

\textbf{\textit{Reflection}:}
In current data videos, \textbf{real-world scenes are typically based on recordings}. 
\review{Q3}\revise{While real-time recording allows for manual alignment but has little room for end-user errors, requiring re-recording in case of mistakes.}
This is both labor-intensive and time-consuming, and existing real-time recording-based methods are all human-driven~\cite{Hall2022, Femi-Gege2024, Zhao2022, Tong2024}. 
\review{Q3}\revise{Post-recording editing alleviates real-time constraints by offering greater flexibility for AI involvement, but end-users cannot alter the recorded content.}

While significant research has focused on data-driven visualization generation in data videos, \textbf{there remains limited support for generating or editing non-data scenes in data video creation}. 
% Although many AIGC tools can assist in creating non-data scenes, and editing tools can aid in post-production, there is still a lack of in-depth research on transition relationships between real-world scenes and visualizations. 
Although many AIGC tools can assist in creating such scenes, and editing tools can be used in post-production, there is a notable gap in research on the transitions between real-world scenes and visualizations. Some studies have explored embedding relationships, but these are typically confined to specific domains or applications~\cite{Chen2022h, Lin2023b, Chen2023d, Chen2022c, Andrews2024}.
The inclusion of real-world scenes and characters can enhance the credibility and authenticity of data videos~\cite{Dasu2024, Mittenentzwei2023a, Sakamoto2022, Sallam2022}, highlighting the need for researching how to seamlessly integrate these elements with visualizations.
Future research could begin by analyzing the visual and semantic transitions between non-data scenes and visualizations in existing data videos, including transitions, embeddings, and cut-ins. Subsequently, systems could be developed that enable users to upload their multimedia content (\eg videos, images, text, \etc), and help them structure a narrative that effectively coordinates real-world scenes with visualizations. 
In cases where necessary materials are missing, the system could further support users by generating or retrieving relevant content.

\review{Q2}\revise{
\textbf{\textit{Summary:}}
In the coordination of real-world scene and visualizations (Tab.~\ref{tab: real-vis}), non-data scenes are primarily based on recordings rather than generated content. Data visualizations can be configured before, during, and after recording, with most enhancements relying on prerecorded videos to augment the visual narrative.
% When involving narratives, various AI assistances are widely adopted to automate labor-intensive tasks like data understanding and animation creation.

}

\subsection{Audio-Enriched Data Videos}
The previous section discussed the construction of animated narratives to convey a complete story. Audio, as a complementary channel, enhances the visual experience by reinforcing the narrative~\cite{clark_dual_1991}. However, introducing audio requires additional consideration of its relationship with visualizations and animations.

\subsubsection{\audio{Audio}}\label{sec:audio}
As shown in Tab.~\ref{tab: audio}, \textbf{audio narration is the most commonly used audio components in data videos}~\cite{Cao2020a}. Many tools allow users to upload or generate scripts that are then converted into audio narration using text-to-speech (TTS) technology~\cite{Ying2023, wonderflow, dataplayer, Latif2022, Andrews2024, Conlen2022}. In recording-based tools, narration typically comes from the presenter’s voice-over~\cite{Hall2022, Zhao2022, Femi-Gege2024}, while some videos are uploaded with built-in original narration~\cite{Shi2023b, Chen2022c}. Additionally, some tools allow uploading and integrating background music to enhance the atmosphere and immersion~\cite{Tang2022}. Sound effects also serve as an important medium for engagement; however, despite existing high-level guidelines~\cite{Xu2022, Herath2023, Rubab2023}, few tools outside of professional software support the integration of sound effects.

\textbf{\textit{Reflection}:}
\textbf{Current tools for audio narration primarily rely on text-to-speech technology}, with some integration of real-time or pre-recorded audio.
\review{Q3}\revise{Text-to-speech technology enables rapid audio file generation but lacks detailed modeling of end-users' nuances (\eg tone and speed), while recording audio offers strong personalization but entails recording complexities.}
% , which can be cumbersome and difficult to modify.
% The existing data video tools that use text-to-speech often use basic audio generation without detailed modeling of nuances like tone and speed.
% Most existing data video tools employing TTS focus on basic audio generation, often lacking detailed modeling of nuances such as tone and speed, which are crucial for conveying emotion and engagement.
Future research could explore audio customization or style transfer from human voices to mimic natural speech.
Additionally, background music (pure or with lyrics) and sound effects are unexplored in data videos, but they are important to create immersive experiences, especially in short-form content like TikTok videos. 
Future work could explore aligning musical elements, such as melody, with narrative structure or generating music or sonification that complements the narrative~\cite{Siu2022}.
% , better aligning emotional tones with narrative content.

\review{Q2}\revise{
\textbf{\textit{Summary:}}
In audio creation (Tab.~\ref{tab: audio}), audio can come from uploaded recordings, live speech, or text-to-speech generation, with text-to-speech being the predominant method.
Narration is the primary use of audio, while music and sound effects are often marginalized.
}

\begin{table*}[t]
% \small
\centering
\caption{Paradigms for audio creation in data video tools (Sec.~\ref{sec:audio}).}
\label{tab: audio}
% \vspace{-10px}

\renewcommand\arraystretch{0.8}
\setlength{\tabcolsep}{3mm}{
\begin{tabular}{lll}
% \rowcolor[HTML]{EFEFEF} 
\toprule
\textbf{Mode} & \textbf{Paradigm} & \textbf{Representative Tools} \\
\midrule

& Video Uploading &  \cite{Chen2022c, Shi2023b}  \\ 
\arrayrulecolor{tablerowcolor} \cmidrule{2-3}
\multirow{-2}{*}{Original} 
& Music Uploading & \cite{Tang2022} \\
% \arrayrulecolor{tablerowcolor} \cmidrule(lr){1-3}
\arrayrulecolor{tablerowcolor} \midrule

\multirow{-1}{*}{Human-Crafted} 
& Speaking and Recording & \cite{Hall2022,Zhao2022,Femi-Gege2024} \\
% \arrayrulecolor{tablerowcolor} \cmidrule(lr){1-3}
\arrayrulecolor{tablerowcolor} \midrule

\multirow{-1}{*}{AI-Generated} 
& Text-to-speech Generation & \cite{ 
Conlen2022,
Latif2022,
Shin2022,
dataplayer,
Ying2023,
wonderflow,
Andrews2024,
NarrativePlayer,
dataplaywright} \\
\arrayrulecolor{tablerowcolor2}\bottomrule
\end{tabular}}
  % \vspace{-10px}
\end{table*}

\begin{table*}[t]
% \small
\centering
\caption{Paradigms for coordinating audio and visualization (Sec.~\ref{sec:AudioVis}) and coordinating audio and  animations (Sec.~\ref{sec:AudioAni}).}
\label{tab: AudioVis}
% \vspace{-10px}

\renewcommand\arraystretch{1.3}
\setlength{\tabcolsep}{1.3mm}{
\begin{tabular}{ccccc}
\toprule
% \hline
\textbf{Audio Source} & \textbf{Paradigm} & \textit{Human-Led}  &  \textit{Mixed-Initiative} & \textit{AI-Led}  \\
% \hline
\midrule
Pre-recorded Audio
& Visual-Audio Alignment
& \cite{premiere, imovie}
& 
& \cite{Shi2023b,Tang2022,Chen2022c}
\\\arrayrulecolor{tablerowcolor}\cmidrule{1-5}

Live Speech
& Real-time Recording
& \cite{Hall2022,Zhao2022,Femi-Gege2024,powerpoint}
& 
& 
\\\arrayrulecolor{tablerowcolor}\cmidrule{1-5}

& Text-Visual Linking
& \cite{wonderflow}
& \cite{Conlen2022,Chen2022c, dataplaywright}
& \cite{ dataplayer, NarrativePlayer}
\\
\multirow{-2}{*}{Narration Text} 
& Narration Generation
& 
& \cite{Andrews2024, Shin2022}
& \cite{Latif2022, Shin2022,Ying2023}
\\
% \hline

\arrayrulecolor{tablerowcolor2}\bottomrule
\end{tabular}}
  % \vspace{-10px}
\end{table*}

\subsubsection{\audio{Audio}---\visual{Visualization}}\label{sec:AudioVis}
This coordination ensures that the audio content aligns with the visual content displayed on the timeline, with audio narration---visualization coordination being predominant.
\textbf{Narration in data videos takes three forms: pre-recorded audio, live speech, and static text converted into audio.} These forms correspond to distinct paradigms, as shown in Tab.~\ref{tab: AudioVis}.
The first is \textit{visual-audio alignment}, commonly used in editing software (\eg Adode Premiere~\cite{premiere}, Apple iMovie~\cite{imovie}, \etc), where pre-recorded audio and animated visual sequences are loaded together. Users repeatedly listen to the audio and watch the visuals, manually ensuring that each frame is precisely aligned. 
The second paradigm is \textit{live speech-based}, where users narrate in real-time while viewing the visual narrative~\cite{Hall2022, Zhao2022, Femi-Gege2024}. Users need to manually control scene transitions and triggering animations, similar to presenting with slides or recording in Microsoft PowerPoint~\cite{powerpoint}.
The third paradigm is \textit{narration-driven} and can be divided into two modes based on the presence of narration text. If narration text exists, static visual elements are linked to text segments, which are then converted into audio with precise timestamps for alignment. For example, WonderFlow~\cite{wonderflow} allows users to link narration segments to visual elements on the interface interactively. Data Player~\cite{dataplayer} further automates text-visual linking using large language models (LLMs), and Narrative Player~\cite{NarrativePlayer} converts long narratives into visual sequences with LLMs.
When narration text is absent, tools like Live Charts~\cite{Ying2023} and Talking Realities~\cite{Latif2022} generate narration text based on insights extracted from visualizations to complete the linking process. 
Additionally, beyond audio narration, SmartShots~\cite{Tang2022} maps multi-modal inputs into shots, corresponding to video segments, and uses an optimization method to match shot durations with the uploaded music rhythm.

\textbf{\textit{Reflection}:}
In existing tools, \textbf{audio is primarily used for narration, with most efforts focused on processing static narration text rather than audio itself}. 
The rise of AI, particularly LLMs, presents new opportunities for narration-driven methods. These approaches either utilize narrative text prepared by users for semantic linking or generate text based on visualizations or data. 
% However, there remains a notable gap in research on converting long narratives into visual sequences, as current text-to-visualization approaches are largely restricted to short texts or commands~\cite{NLISurvey}. 
However, the interplay between visualizations and musical elements or sound effects has yet to be thoroughly explored.

\review{Q3}\revise{Furthermore, \textbf{users’ preferences for handling narration are also influenced by the narration source}. Narration text can be categorized into two types: original content created by the author and content created by others, with the author performing secondary tasks like enhancing visuals. For original content, users often prefer to lead the process by uploading their narration text or providing live speech. As LLMs continue to advance, an increasing number of users may seek assistance from these models for crafting narration~\cite{Ying2023}. In contrast, for content created by others, where the task pattern is more defined, AI assistance is already more prevalent~\cite{NarrativePlayer}.}

\review{Q2}\revise{
\textbf{\textit{Summary:}}
In the coordination of audio and visualizations (Tab.~\ref{tab: AudioVis}), temporal alignment between the two aspects is crucial. Current efforts primarily focused on processing static narrative text to establish synchronized text-visual connections with AI support, compared with manually aligning visuals and audio content.

% visualizations with new paradigm design and AI technologies.
% workflow are streamlined by both interaction paradigm design and AI technologies.
}

\subsubsection{\audio{Audio}---\motion{Motion}}\label{sec:AudioAni}
Compared to coordinating audio with visualizations, enabling narration-animation interplay focuses on aligning audio narrative content with animation triggers and durations~\cite{Borgo2022}. 
As shown in Tab.~\ref{tab: AudioVis}, \textbf{narration-driven paradigms remain mainstream similar to audio---visualization coordination}, and most of them are built on text-visual links. For instance, WonderFlow~\cite{wonderflow} allows users to select an animation preset for each text-visual link, while Data Player~\cite{dataplayer} uses constraint programming to recommend a suitable animation sequence for all text-visual links. 
This typically assumes the presence of both the visualization and narration text.
In cases where neither exists, users generally input narration text segments, which are further parsed for automatic animated visualization generation. This scenario often involves an uploaded video. For example, in Sporthesia~\cite{Chen2022c}, users first select the sports video segment they wish to enhance, input the corresponding commentary text, and the system automatically converts the commentary into audio and generates the matching animated visualization. AiCommentator~\cite{Andrews2024} allows users to input specific questions at particular moments (\eg ``\textit{Which team is performing better this season?}''). The system then automatically extracts relevant events and objects, generates animated visualizations, and uses LLMs to produce the commentary. 
Furthermore, Roslingifier~\cite{Shin2022} takes an animation-driven approach, allowing users to add narration to specific events after an animated narrative is automatically generated. In addition, manual coordination based on recording is similar to audio---visualization coordination~\cite{Hall2022, Zhao2022, Femi-Gege2024}.

\textbf{\textit{Reflection}:}
The coordination of audio and motion aligns with the findings in Sec.~\ref{sec:AudioVis}, as motion effects need to be applied to visuals. Beyond this, \textbf{an unexplored paradigm is animation-driven coordination}.
Animation can convey specific information, such as data insights~\cite{Shi2021a} or emotions~\cite{Lan2023, Xie2023, Lan2022}, and users could select preferred animations, which would then automatically adapt to the visualization and generate narration text and music, similar to existing example-driven methods~\cite{galvis, Xie2023a}.

\review{Q2}\revise{
\textbf{\textit{Summary:}}
In the coordination of audio and motions (Tab.~\ref{tab: AudioVis}), it mirrors similar paradigms seen in audio-visualization coordination, typically accomplished through manual alignment or AI recommendations based on established text-visual links.
}

% Moreover, developing animation-driven audio narration generation could enhance storytelling processes, providing a more immersive experience for viewers. The collaboration between human creativity and AI-driven enhancements is crucial for advancing narrative structures in data videos.

\section{Discussion}
\review{Q6}\revise{
% The previous section analyzed and reflected on the design paradigms of each component and their coordination. 
This section presents the identified gaps along two dimensions in our framework, as well as the limitations and future work. 
Since this paper focuses on technical implementation, we will first discuss gaps in the second dimension (\textit{how to support the creation and coordination of data video components}). Moreover, to develop more effective design paradigms, a deeper understanding of data video components and their coordination is needed. So then we will discuss relevant gaps in the first dimension (\textit{what data video components to create and coordinate}) and how our framework may help address the gaps.
}
% (\textit{what data video components to create and coordinate}, and \textit{how to support the creation and coordination})
% This section will take a holistic view of data video creation, identifying remaining gaps in the whole process. We also shed light on how to bridge the gaps at each stage, including empirical studies, component creation and coordination, interaction design, evaluation, and applications. 
% to inspire future research.

\revise{
\subsection{How to Support the Creation and Coordination of Data Video Components}
While existing design paradigms for each component in Sec.~\ref{sec:tools} can effectively assist users in authoring data videos, some gaps remain for future exploration.
}

\begin{figure}[t]
  \centering
    \includegraphics[width=0.9\linewidth]{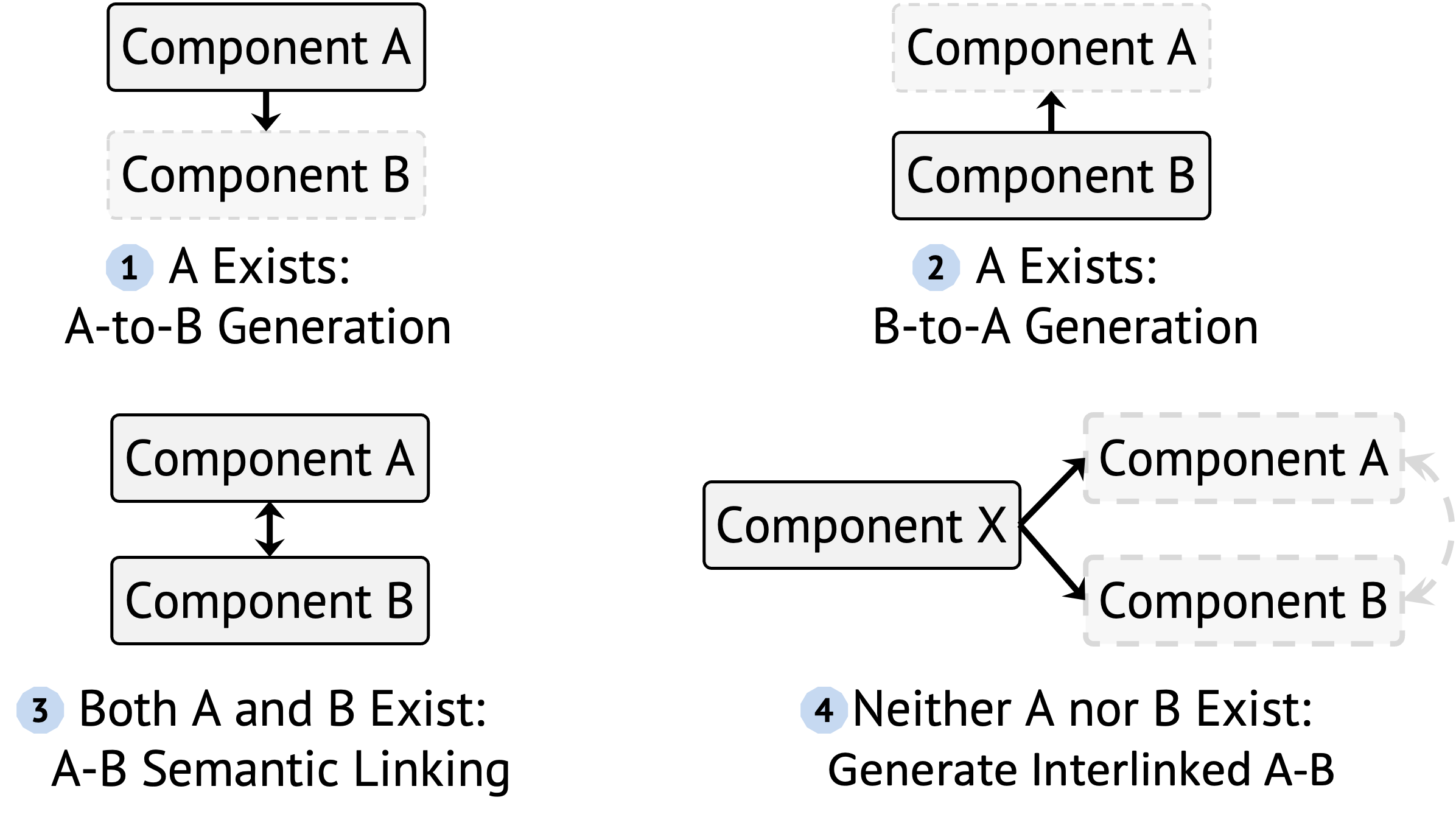}
% \vspace{-10px}
    \caption{Coordination patterns of two data video components, depending on the presence of the two components.}
\label{fig:Coordination}
  % \vspace{-10px}
\end{figure}

% \subsection{Data Video Component Creation and Coordination}
% \textbf{Gap 5: Greater diversity of components and their coordination lead to better expressivity potential, but current tools cannot fully support all aspects.}
% , as well as our reflections. 
% Here, we offer some additional high-level discussion. 
% Data video components are fundamental to message delivery, and their effective coordination is crucial to production. 
% Data video components are foundational to conveying messages, and their coordination is a critical aspect of video production. 
% For the two types of components discussed before, the existing coordination patterns generally fall into four categories, as depicted in Fig.~\ref{fig:Coordination}, depending on the presence of the two components. 
% Sec.~\ref{sec:tools} provides a detailed analysis of current research on data video components and their coordination. The coordination patterns for two component types generally fall into four categories, as shown in Fig.~\ref{fig:Coordination}, based on their presence. 
\review{Q10}\revise{
\textbf{Gap 1: A wider range of data video components and their coordination offers larger design space, but current tools lack comprehensive support for all aspects.}}
Sec. \ref{sec:tools} analyzes current research on the creation and coordination of data video components, and the coordination of two components generally fits into four categories based on their presence, as illustrated in Fig. \ref{fig:Coordination}.
While data video components and their coordination present a vast design space with significant expressive potential, existing tools fail to fully support this diversity, limiting the ability to expand the creative possibilities of data videos. 
% Data video components and their coordination form a large space, but current tools cannot fully support all aspects to expand the expressivity potential of the output data videos. 
Future tool development efforts could integrate and coordinate a broader range of components, such as real-world scenes and pictographs in visuals, and music and sound effects in audio, as discussed in Sec.~\ref{sec:real-vis} and Sec.~\ref{sec:audio}.
Achieving this will require integrating advancements in computer vision, graphics, and audio processing technologies. 
Additionally, the development of a robust representation language to unify and manage these diverse components within a single system, as well as the creation of new design paradigms, will be essential.
% alongside a robust representation language to manage all components within a unified system and the design of new paradigms.
However, it is also important to note that more diverse components do not necessarily equate to more expressive data video output. Creators should understand each component's purpose, select appropriate combinations within a specific context, and employ them effectively, which will be further discussed in Gaps 6 and 7.
% Achieving expressiveness requires a deep understanding of each component's functionality and intended use, identifying the most suitable combination of components within a specific context, and effectively designing and utilizing these components, as outlined in Gap 1 and Gap 2.

\textbf{Gap 2: Current tools mainly focus on reducing user learnability, but rarely expand expressivity.}
\review{Q6}\revise{According to our analysis, many current tools focus on user-friendly paradigms and automation to reduce learnability when involving more components vertically~\cite{Grossman2009}, showcasing their expressivity through limited examples.} However, expanding expressivity horizontally is often deferred to future engineering efforts. 
For instance, in template-based systems, the expressivity of data videos is constrained by the diversity of templates, which are time-consuming to extend~\cite{wonderflow, Amini2017}.
A key future direction is to explore paradigms to rapidly enhance expressivity while ensuring ease of learning. This could involve efficiently creating templates from real-world examples, reverse-engineering data videos to extract features, and adapting them to current tool frameworks.
% A key future direction is to explore rapid expressivity expansion while maintaining learnability. This could involve efficient templatization of real-world example galleries, reverse-engineering of data videos to extract features, and automatically adapting them to existing tool frameworks.
In addition, from a tool perspective, components can be categorized into two types: user-uploaded and tool-generated. 
For user-uploaded components, tools should provide advanced analytical capabilities in understanding and effectively integrating these elements into final data videos.
For tool-generated components, the focus should be on enhancing the generation process to align closely with user intent.
% User-uploaded components require advanced understanding to support design, while tool-generated components need strong capabilities to fulfill user intent. 
% Future developments should leverage cutting-edge AI models to enhance both understanding and generation.
% Moreover, a rising direction is building multi-agent systems to simulate each role in data video creation and fully automate the process~\cite{Shen2024a}.

% \subsection{Interaction Design}

% \noindent
% \textbf{Fine-Grained Intent Modeling.} 
\textbf{Gap 3: Tools can only handle a defined set of user intents, but users may have more diverse intentions.}
The goal of interaction is to enhance communication between users and tools, allowing for the exchange of intents and feedback. 
% Given the complexity of tasks, 
User intents can vary widely, from abstract goals (\eg ``\textit{create a magical video}'') to specific tasks (\eg ``\textit{highlight the blue bar}''), forming a hierarchical structure. 
Different user roles, such as novices and experts, also have unique intents and requirements.
\review{Q6}\revise{Currently, our understanding of user intents in data video creation is fragmented in various tools and lacks a unified model.}
Such a model could bridge varied user intents with specific tool functions, enabling users to articulate their objectives more effectively and aiding the system in accurate interpretation.
% Currently, the understanding of user intents in data video creation is fragmented and lacks a comprehensive model. Developing a detailed model of user intent is essential for mapping these varied intents to corresponding actions within the tool. This model would enable users to better articulate their intents at different levels, while allowing the system to more accurately interpret and respond, leading to more effective interaction design.
% \noindent \textbf{Interaction Preferences for Each Task.} 
% In complex data video creation tasks, the more freedom a tool grants the user, the greater the demand on the tool's ability to parse user intents and generate corresponding components. 
Additionally, granting users more freedom increases the need for tools to understand their intents and generate suitable components.
Existing tools often limit user freedom through strategies like predefined templates~\cite{Amini2017, wonderflow}, rules~\cite{Li2023b, Wang2021d}, or syntaxes~\cite{Ge2020, Zong2022}. Although effective, users often face a learning curve and may find it hard to understand the system's capabilities and best inputs. 
\review{Q6}\revise{Certain tools also limit the editing of specific components (\ie Original mode in Fig.~\ref{fig:learnability}) or automate the process entirely (\ie AI-Led mode in Fig.~\ref{fig:learnability}), which might not meet users' customization requirements.}
% Some tools also restrict certain components from being edited or fully automate the process, which may not satisfy users' customization needs.
Therefore, in the future, we could comprehensively summarize user preferences in communicating component-specific tasks to tools, especially regarding the balance between human and AI involvement and favored collaboration methods, to develop tailored interaction paradigms~\cite{Li2023a}.
% This understanding can help balance user intent with the tool’s complexity and implementation challenges. Additionally, gathering these preferences can vary depending on the scenario, whether optimizing tools for professionals or making them accessible to novices.
Finally, interaction design should align with established patterns~\cite{nielsen1999designing}. When introducing new interaction paradigms, it is important to consider how to educate users on their use and how to persuade them to adopt new workflows over existing ones.

\textbf{Gap 4: LLMs are increasingly applied in data video creation, but their reliability remains difficult to ensure.}
\review{Q6}\revise{The rise of LLMs has significantly accelerated tool development and facilitated research on new design paradigms for different data video components, especially within mixed-initiative and AI-led transformation modes~\cite{dataplayer, dataplaywright, Andrews2024, Shen2024a, Ying2023}. However, it also raises concerns about reliability and what we should do when encountering errors.} 
Enhancing the robustness of AI-driven tools requires not just advanced error correction but also greater transparency in AI decision-making processes. 
% Future research should focus on enhancing the reliability of AI-driven tools by developing advanced error correction mechanisms and increasing transparency in AI decision-making. 
Integrating human-in-the-loop approaches could provide a safeguard, allowing human oversight to intervene when AI outputs are unreliable. 
Moreover, hybrid evaluation frameworks that blend AI efficiency with human judgment are essential to ensure tools remain not only powerful but also trustworthy, especially in complex real-world applications.
% Additionally, developing hybrid evaluation frameworks that combine AI strengths with human expertise can ensure that tools are not only powerful but also trustworthy and adaptable to real-world conditions.

% \noindent
% \textbf{Reliability of AI Models.} 
% The rise of large AI models has significantly accelerated tool development but also raises concerns about reliability and handling errors. Current practices include model interpretability, error detection mechanisms, and robustness testing, yet there is room for improvement. Future research should focus on enhancing the reliability of AI-driven tools through better error correction strategies and increased transparency in AI decision-making. Integrating human-in-the-loop approaches could provide a safeguard, allowing human oversight to intervene when AI outputs are unreliable. Additionally, developing hybrid evaluation frameworks that combine AI strengths with human expertise can ensure that tools are not only powerful but also trustworthy and adaptable to real-world conditions.

% \subsection{Evaluation}
% \noindent
% \textbf{Establish An Evaluation Framework.} 
\textbf{Gap 5: Data video tasks are complex, but a comprehensive evaluation framework is lacking.}
\review{Q6}\revise{
Data video tools, involving diverse components, have different evaluation focuses and strategies.
}
The evaluation typically involves two key aspects: the quality of the generated videos and the effectiveness of the design paradigms. 
While criteria for data videos (\eg comprehension, memorability, and engagement) and for tools (\eg usability, extensibility, and expressivity) have been identified, they often remain general and are typically assessed through human-centric methods like user studies and expert interviews~\cite{Riche2018}. 
% This reliance on subjective evaluation is understandable given the diverse nature of data storytelling, where individual preferences are hard to model. 
Recent efforts have begun to introduce quantitative approaches, such as modeling story transitions~\cite{Shi2021a}, aligning music with video content~\cite{Tang2022}, counting interactions~\cite{wonderflow}, and predicting design outcomes~\cite{Wang2021d}. 
These are promising steps, but there is a clear need for a more comprehensive, controlled, and quantifiable evaluation framework. 
Insights from cognitive science and psychology could help inform the development of such a framework, combining qualitative and quantitative measures. 
Additionally, the advancement of LLMs could enhance the evaluation process at various stages, particularly in areas that are hard to quantify. For example, AI models could simulate diverse scenarios and target audiences, enabling more personalized and context-aware evaluations.

\subsection{What Data Video Components to Create and Coordinate}
\review{Q6}\revise{
% Sec.~\ref{sec:tools} provides a detailed discussion on the creation and coordination of data video components based on our decomposition (Sec.~\ref{sec:components}). 
To develop more effective and targeted design paradigms and tools, a deeper and more comprehensive understanding of the data video components and their interplay in combination is essential.}
% \subsection{Empirical Research for Comprehensive Data Video Understanding}

% \textbf{Comprehensive Exploration of Data Videos.}
\textbf{Gap 6: Data videos exhibit diverse media forms, but lack a refined classification system and clear characteristic modeling.}
\review{Q6}\revise{Since the emergence of a broad definition for data videos~\cite{Segel2010, Amini2015}, they have been observed in diverse contexts, such as social media, art exhibitions, and science communication, with different data video component combinations.
% they have manifested in various forms across diverse scenarios like social media, art exhibitions, and science communication. 
Despite their popularity, when building our framework, we found that there is no comprehensive classification that captures their diverse characteristics.
Our framework serves as a starting point for technical implementation. In the future, it could be expanded to refine the purpose of each component within data videos (\eg the purpose of audio narration could include double encoding of visual information, highlighting specific visuals, enhancing visuals, complementing visuals, \etc), thus deducing a comprehensive space to better grasp their characteristics and inspire future tool development.
}
Furthermore, there are some other perspectives that deserve exploration, such as narrative intent (\eg showcasing data, explaining visualizations, narrating stories, \etc), communication goals (\eg description, hypothesis, persuasion, \etc), data presentation formats (\eg charts, pictographs, encoded in real-world scenes, \etc), and audience engagement (\eg passive viewing, interactive exploration, \etc). 
Such refined modeling can facilitate a more targeted paradigm design for different data videos.
Additionally, while many aspects of data videos have been studied, there are still some unexplored design spaces, such as the use of music and sound effects, and transitions between real-world scenes and visualizations, as noted in Sec.~\ref{sec:tools}. 
% Despite their popularity, there is still a lack of comprehensive classification that summarizes the diverse characteristics of data videos. 
% This classification could consider factors such as narrative intent, communication goals, data presentation formats, and modes of audience engagement. 
% \review{Q6}\revise{Our current framework's component decomposition mainly focuses on the component forms. }

\textbf{Gap 7: Data videos are seen as advantageous, but their ideal contexts remain unclear.}
\review{Q6}\revise{Each data video component plays a distinct role in engaging the audience, and their combinations lead to varying interplays, applied in various contexts. Nonetheless, the ideal contexts for different component combinations and usage remain obscured.}
While many studies highlight the advantages of data videos~\cite{Amini2018a, Robertson2008, Rodrigues2024}, they also possess some inherent limitations, such as requiring viewers to follow a rigid narrative pace and absorb much information in a short time~\cite{Riche2018}. 
We cannot always expect an ``easy win'' of data videos over simpler forms (\eg static visuals)~\cite{Amini2018a, Kong2019}; there are numerous trade-offs in effectiveness, production cost, and viewer engagement. 
Therefore, more research is needed to assess the overall effectiveness of data videos and determine the best contexts for utilizing different components.
Such enhanced understanding of data videos can potentially lead to better paradigm design and tool development.
% A more comprehensive empirical investigation will enhance our understanding of data videos and potentially lead to better paradigm design and tool development.
\review{Q6}\revise{
Our framework could initiate a bottom-up analysis of each component's role and their interplay, ultimately deriving the appropriate contexts for different component combinations.
}

% improve their technical implementation.
% enrich the knowledge base of data videos and provide further guidance for technical implementation.

% \noindent
% \textbf{Bridging the Gap Between Empirical Research and Creation Tools.}
% \textbf{Gap 3: There are numerous empirical studies, but they are rarely directly applied to tool development.}
% , as discussed in Sec.~\ref{sec:related}
\textbf{Gap 8: Many empirical studies exist, but few are directly applied to paradigm design and tool development.}
% Existing empirical studies have provided various guidelines for data videos, providing rich knowledge for the design of paradigms.
% However, these guidelines are rarely directly applied in authoring tools~\cite{Chen2022}, \review{Q6}\revise{as discussed in Sec.~\ref{sec:VisAni}.}
\review{Q6}\revise{Complex coordinated components in data videos have been explored through many empirical studies, yielding rich knowledge (\eg guidelines and design spaces) from various perspectives. Such knowledge can contribute to the design of authoring paradigms.
For example, Chen \etal~\cite{Chen2022} considered design spaces as an initial stage of tool automation, enabling creators to engage in more objective design processes and facilitating automated tools' understanding of creators.
However, these knowledge are rarely directly applied in authoring tools, as discussed in Sec.~\ref{sec:VisAni}.}
Several factors may contribute to this gap: 
first, empirical studies often focus on design, while tools emphasize measurable and computable authoring processes; 
second, the output of empirical research is typically general, whereas tools are context-specific and require exploring a distinct design space; 
third, many effects explored in empirical studies rely on human design to achieve within existing components (\eg emotional impact~\cite{Lan2023, Lan2022}, narrative structure~\cite{Lan2021a, Yang2022a, Wei2024}, and cinematic effects\cite{Xu2022, Xu2023b}), whereas tools tend to focus on explicitly implementable functions; 
fourth, the abundance and dispersion of empirical guidelines make it difficult for creators to identify the most relevant ones.
\review{Q6}\revise{To bridge this gap, the community could maintain a unified knowledge base for data videos, categorizing the existing guidelines, and potentially starting with our framework (by each component and coordination).} 
Additionally, a computable representation framework could be established, mapping high-level guidelines to low-level constraints~\cite{dataplayer}, and also designing interfaces that integrate new knowledge effectively.

% \subsection{Expending the Application Scope}
\textbf{Gap 9: General data video research is flourishing, but its penetration in domain applications remain limited.}
While general research on data videos is important, many fields either overlook or fail to utilize them effectively. 
\review{Q6}\revise{According to the analysis in Sec.~\ref{sec:tools}, we found that research on the application of data videos in specific domains often involves real-world scenes (Sec.~\ref{sec:real-vis}), and current domain-specific research primarily concentrates on sports~\cite{Chen2022c,Chen2023d,Chen2022h,Lin2023b,Lin2023a, Yao2024} and health~\cite{Sakamoto2022, Sallam2022}, both empirically and in tool development.}
% Future efforts should deepen this integration.
Therefore, a crucial direction is to further integrate data video techniques within specific domains and explore domain-specific design paradigms. 
One strategy is to extend the use of videos in fields where they are already employed to enhance communication, such as dynamic code presentations~\cite{NotePlayer}, news reviving~\cite{NarrativePlayer}, and video-based programming tutorials~\cite{Chi2022}. Data-driven modeling can build on these contexts. %news reviving~\cite{Wang2024f}
Another strategy is to expand into fields well-established in Visual Analytics Science and Technology (VAST) research~\cite{munzner2014visualization}, integrating storytelling approaches to bridge data analysis and insight communication.
% By leveraging VAST methodologies and integrating storytelling components, we can better bridge the gap between data analysis and insight communication.
% Furthermore, developing domain-specific frameworks that capture the unique characteristics of each field can help data videos adapt more effectively.
In addition, while data videos primarily serve communication purposes, their potential can extend beyond this. 
% For example, data videos and other storytelling forms can enhance the data analysis process by presenting intermediate results in a narrative format, aiding analysts in decision-making. 
For example, they can aid in the data analysis process by presenting results in a narrative format, aiding decision-making. 
Data videos, as dynamic presentations, can also improve explainability, such as in the interaction and feedback loop between users and LLMs~\cite{xie2024waitgpt}. 
\review{Q6}\revise{These interesting applications also require the design of new paradigms. The paradigms summarized in this paper could serve as effective references.}

\subsection{Limitation and Future Work}
Our research has several limitations. 

\review{Q5}\revise{\textbf{Lack comprehensive analysis of commercial software.}
% We focused primarily on analyzing academic tools, without conducting a comprehensive survey of commercial software. On the one hand, it is challenging to capture the full scope of commercial tools; on the other, these tools evolve continuously, making it difficult to summarize paradigms and compare them with other tools. We hope that future research will continuously integrate commercial software into our analysis framework.
% We primarily focused on analyzing academic tools, without conducting a comprehensive survey of commercial software. Due to their extensive and varied nature, capturing the complete scope of commercial tools poses a challenge.
% While commercial software prioritizes practicality, and academic software focuses on innovation, both realms influence and propel each other. Additionally, commercial software can benefit from our decomposition and analysis framework.
Our emphasis has been on academic tools, and as such, we have not thoroughly surveyed commercial software due to its diverse and extensive nature.
Commercial software differs in emphasis from academic software: the former prioritizes practicality, while the latter prioritizes innovation, yet they mutually influence each other. 
Additionally, commercial software can also benefit from our framework for component decomposition and analysis.
In Sec.~\ref{sec:tools}, we also mentioned some typical commercial software design paradigms, such as Flourish~\cite{flourish} that adopts an AI-led template-driven paradigm for motion creation, and PowerPoint~\cite{powerpoint} that features human-led real-time audio and visualization coordination.
Future research could consistently integrate commercial software into our analytical framework. These tools can be broadly categorized into data-driven tools (\eg Flourish~\cite{flourish} and Gapminder~\cite{Gapminder}) and general tools (\eg Adobe After Effects~\cite{aftereffects}, iMovie~\cite{imovie}, and PowerPoint~\cite{powerpoint}). 
Data-driven tools align closely with the academic tools discussed in this paper. 
However, general tools offer a wide range of features and evolve continuously, often lacking explicit modules tailored for data video authoring. Therefore, a more thorough examination is necessary to identify relevant modules for data video authoring before incorporating them into the analysis.
% Moreover, our analysis in Tab.~\ref{tab:tools} was primarily vertical and organized by components, as these tools' main contributions often lie within a few key components, yet it lacks comprehensive horizontal discussions at the overall tool level. 
% Given that each tool comprises multiple components, its main contributions often lie within a few key components.
% Commercial software typically offers a wide range of features and evolves continuously. 
% A horizontal discussion can help better grasp its composition, module interactions, and inter-software distinctions.
% In the future, we envision incorporating a horizontal analysis of overall tools within our existing vertical framework
% In the future, expanding the horizontal dimension regarding the overall tools on top of our vertical analytical framework can be considered.
% A horizontal discussion tends to obscure the design paradigms, while a vertical discussion allows for deeper insights.
% Second, our analysis of Tab.~\ref{tab:tools} is primarily vertical, organized by components, and lacks a detailed horizontal discussion at the tool level. Since each tool comprises multiple components, its main contribution is often concentrated in a few key components. A horizontal discussion tends to obscure the design paradigms, while a vertical discussion allows for deeper insights.
}

\review{Q7}\revise{
\textbf{Relatively limited datasets.}
Our analysis is based on 46 papers, selected according to the criteria discussed in Sec.~\ref{sec:Landscape}, to explore the design paradigms of data videos. 
This straightforward dataset is tightly focused on data videos and thus relatively small, it may not fully capture the breadth of design paradigms, particularly in more complex scenarios. However, it provides a useful starting point. As this topic is still in its early stages and gaining increasing attention, as shown Fig.~\ref{fig:overview}, future work could consistently integrate the latest work into our framework for further analysis.
% Additionally, expanding the dataset to include more diverse and multifaceted sources, such as commercial software or general video creation practices, could help identify transferable paradigms, validate our findings, and deepen understanding of the field.
Furthermore, we can enhance our studies by gathering diverse and multifaceted datasets. For example, we can conduct a more thorough examination of commercial software, as discussed before. 
Investigating general video creation practices will also allow us to analyze their connections and distinctions with data videos, validate our findings, and identify transferable paradigms.
}

% \textbf{Lack end-user and audience perspective in the framework.}

\review{Q3}\revise{\textbf{More fine-grained and multifaceted framework.}}
Our annotation of tools in the framework (Tab.~\ref{tab:tools}) could have been more fine-grained, such as at the specific component level discussed in Sec.~\ref{sec:components}, rather than the clustered component types in Sec.~\ref{sec:expressivity}. However, we did not adopt this approach because such annotation would be too detailed, hindering the generalization and discussion of design paradigms. Moreover, some sub-tasks are not unique to data video creation.
\review{Q3}\revise{Furthermore, our framework could be more diverse. Currently, it focuses on the creation and coordination of data video components, without fully considering the impact on end-users and audience perception~\cite{Lee2015}. Although the reflection parts in Sec.~\ref{sec:tools} provide some relevant discussions, future work could expand the framework for a more comprehensive analysis of how each component and coordination affect end-users and audiences.}

% \sutt{One promising approach for supporting more accessible and appealing human-data interactions is to present data in media, entertainment and cultural forms people are already familiar with and enjoy.}

% \sutt{We found that those using a small multiples design (static) consistently completed tasks in less time, albeit with slightly less confidence than those using an animated design. The accuracy results were more task-dependent}

\section{Conclusion}
In this paper, we proposed a framework to systematically understand existing data video creation tools, reflecting on key design paradigms from the perspectives of components and coordination. We also discussed the remaining gaps in data video production and highlighted potential future directions. In the future, we plan to continue tracking the latest research developments and update our corpus accordingly. We hope that the presented framework and insights will raise awareness of this topic and inspire further interesting work. Additionally, we will follow up on the identified gaps, especially by designing new paradigms and developing tools. Moreover, we plan to extend the gained insights to other forms of storytelling, such as scrollytelling and slides.

\begin{acks}
The authors wish to thank Liwenhan Xie and Lin Gao for their valuable suggestions that have enhanced the paper's quality.
This work is partially supported by Hong Kong RGC GRF grant (16210722).
\end{acks}

\bibliographystyle{ACM-Reference-Format}

\bibliography{main}

% \clearpage
% \iffalse
% \appendix
% % \input{sections/VIS_appendices.tex}
% \fi

\end{document}